# Predictive biomarker graphical approach (PRIME) for Precision medicine

Gina D'Angelo[a], Xiaowen Tian[a], Chuyu Deng[b], Xian Zhou[b]

[a]Oncology Statistical Innovation, Oncology R&D, AstraZeneca, Gaithersburg, Maryland

[b]Oncology Biometrics, Oncology R&D, AstraZeneca, Gaithersburg, Maryland

*Corresponding author: Gina D'Angelo, AstraZeneca, One MedImmune Way, Gaithersburg, MD 20878;

email: gina.dangelo@astrazeneca.com

**Abstract**

Precision medicine is an evolving area in the medical field and rely on biomarkers to make patient enrichment decisions, thereby providing drug development direction. A traditional statistical approach is to find the cut-off that leads to the minimum p-value of the interaction between the biomarker dichotomized at that cut-off and treatment.  Such an approach does not incorporate clinical significance and the biomarker is not evaluated on a continuous scale.  We are proposing to evaluate the biomarker in a continuous manner from a predicted risk standpoint, based on the model that includes the interaction between the biomarker and treatment. The predicted risk can be graphically displayed to explain the relationship between the outcome and biomarker, whereby suggesting a cut-off for biomarker positive/negative groups. We adapt the TreatmentSelection approach and extend it to account for covariates via G-computation. Other features include biomarker comparisons using net gain summary measures and calibration to assess the model fit. The PRIME (Predictive biomarker graphical approach) approach is flexible in the type of outcome and covariates considered. A R package is available and examples will be demonstrated.

# 1  Introduction

Precision medicine is an active area. This involves identifying a biomarker and its cut-off to find the right patient population for enrichment or stratification. In the pharmaceutical industry understanding which patient population will benefit from treatment is crucial. This requires statistical methods that can guide strategic decisions being made to drive the portfolio and regulatory interactions is key. Some examples of predictive biomarkers are HER2 for breast cancer, BRAF mutation for melanoma patients, and CFTR mutations for cystic fibrosis patients that have led to targeted therapies (Abrahams 2006; Jorgensen 2019). Biomarkers are usually measured on a continuous scale. The objective in clinical trials is to find the "enriched population", and to achieve this an "optimal" cutoff needs to be identified.

Traditional approaches to finding a cut-off may involve doing a "by" or a "subgroup" analysis where a model is analyzed by the different categories of the biomarker. However this includes a subset of the data and is a loss of efficiency and can lead to bias (Faraggi and Simon 1996). Another approach commonly used evaluates the model with treatment and a dichotomized biomarker and identifies the cut-off that leads to the smallest p-value, known as the minimum p-value approach. It has been demonstrated that finding the cut-off using a minimum p-value approach can lead to an inflated type-I error and overestimation (Faraggi and Simon 1996; Royston, Altman, and Sauerbrei 2006; Woo and Kim 2020). Faraggi et al (Faraggi and Simon 1996) have shown benefit to adjusting for the type-I error and cross validation for the minimum p-value approach. Some other approaches that are focused on finding a cut-off include the Bayesian hierarchical model (Chen, Jiang, and Tu 2014) approach and minimum p-value approaches and their extensions ((Faraggi and Simon 1996), (Woo and Kim 2020)). However, there is benefit to understanding if the biomarker (as continuous) is considered predictive, then find the cut-off that may lead to the right subpopulation while analyzing the biomarker as continuous.

A more informed approach is to analyze the biomarker as continuous. Royston et al. (Royston, Altman, and Sauerbrei 2006) and Austin et al (Austin and Brunner 2004) suggest to analyze the data as continuous and that dichotomization can lead to a loss of power, and using all the data available is preferred.  There has been an increase on precision medicine and methods to evaluate the biomarker as continuous variables. There have been much research that focuses on the descriptive relationship and modelling the biomarker and outcome (Bonetti and Gelber 2004; Royston and Sauerbrei 2004; Ware 2006; Wang et al. 2007; Cai et al. 2011; Claggett et al. 2011; Werft, Benner, and Kopp-Schneider 2012; Zhao et al. 2013). While our objectives are similar, our scope is broader, as we include suggestions for cut-off values, comparisons of biomarkers, and assessment of model performance.

A more informative and improved approach to identify the biomarker as "predictive" that uses all the data in the model is a model that includes treatment and biomarker interaction. This approach assesses if the magnitude and/or direction of the relationship between efficacy and the biomarker differs by treatment, while keeping biomarker in its original form, e.g. continuous. The next step involves graphing the predicted outcome as predicted risk versus the biomarker by treatment from the model with interaction. Janes et al.  (Janes et al. 2011; Janes, Pepe, and Huang 2014) developed this approach and provided guidance on how to find a cut-off and demonstrate treatment benefits.  We extend this approach to adjust for covariates while considering various types of outcomes. In clinical trials we often adjust for prognostic factors, and decisions often rely on models that include these variables. Often times there is a need to adjust for some baseline covariates, requiring methods that account for them while averaging out their effect. Particularly in the context of randomized trials, we often have to adjust for covariates to account for their impact, while also generalizing the cut-off recommendation to the broader population.

We refer to this approach as PRIME (Predictive biomarker graphical approach). This approach generalizes the treatment selection approach (Janes et al. 2014) to identify predictive biomarkers and its cut-off where treatment benefit begins. Our aim is to look for a predictive biomarker where the relationship between the treatment and efficacy differs by biomarker subgroups.  There could be a magnitude difference only (quantitative interaction) or treatment switching by biomarker subgroup (qualitative interaction) (Polley et al. 2013; Roth and Simon 2018). We are primarily focused on qualitative biomarkers where the risk curves, as a function of the biomarker values, for the treatment and control groups cross, limiting the focus to finding a cut-off in this scenario. We also suggest additional cut-offs in addition to the biomarker value where the risk curves cross. This approach has been generalized to adjust for covariates and also for three types of outcomes (continuous, binary, and time-to-event). G-computation has been used to marginalize covariates, also known as averaging out the covariates. We also propose a way to use net benefit summary measures to compare biomarkers and biomarkers with various cut-offs. Further we assess the model fit with calibration. We also propose an extension to handle the summary measure and calibration while adjusting for covariates.

Section 2 will discuss the PRIME method, Bayesian hierarchical model approach and minimum p-value approach. Section 3 presents simulation studies to understand the operating characteristics of the PRIME proposed approach compared to the Bayesian hierarchical model approach and minimum p-value approaches. Section 4 describes an example, and we close with a discussion.

## 2  Method

### 2.1  Biomarker and cut-off

Typically, one will assess if a biomarker is predictive by only evaluating whether the interaction between the dichotomized biomarker and treatment is statistically significant. The cut-off to be selected would be the cut-off that leads to the minimal p-value. This is not the best approach as the biomarker relationship has not been evaluated on a continuum to observe the overall relationship, what changes and when. Also, many tests are conducted that lead to correlated results and multiple testing, which can cause an inflated type-I error. An improved approach that lends itself to finding a clinically meaningful biomarker is to visualize the relationship to see the clinical magnitude of the relationship between the outcome and biomarker by treatment and to observe these relationships along the distribution of the biomarker, thereby suggesting when the treatment of interest has better efficacy than the standard of care arm (SOC). It is also suggested to first assess if the interaction between the biomarker and treatment is statistically significant, when the biomarker is continuous.

This approach generalizes the treatment selection approach (Janes et al. 2014) to identify predictive biomarkers and its cut-off where treatment benefit begins, and it has been adapted to adjust for covariates and also for three types of outcomes (continuous, binary, and time-to-event). G-computation has been used to marginalize the covariates, also known as averaging out the covariates. The focus is on predicting the marginalized risk so the probability of the worse outcome will have a higher value or the probability of the outcome has to be flipped to indicate worsening event if the higher outcome value indicates better outcome such as in survival. There is a clinical interpretation to the plots when we average over covariates, as the teams prefer not to analyze the relationships based on a single covariate value.

## 2.2 PRIME approach

This PRIME approach fits a model to the data and uses prediction to evaluate the relationship between the outcome and biomarker by treatment while accounting for other covariates. The model will contain treatment, biomarker as continuous, interaction between treatment and biomarker, and covariates that may be referred to as prognostic factors. In this manuscript, we assume a linear relationship between the outcome and biomarker, and will discuss possible alternative approaches later. The models considered here include the generalized linear model (GLM) and the Cox proportional hazards model so we will assume linearity, homogeneous variance, proportional hazards. Alternative options will be addressed in the discussion.

The PRIME approach has a number of features that will be useful to guide the project team in visualizing the relationship between the outcome and biomarker by treatment. It helps to determine if the treatment has improved efficacy over the control arm and where the efficacy magnitude becomes larger, assess when there is treatment benefit for the treatment arm and suggests a cut-off where this benefit occurs, and compare biomarkers and cut-off values. This all can be evaluated while adjusting for covariates and is done using marginalization over the covariates. Furthermore, we use model-based derived confidence intervals for the predicted risk, which differs from the Treatment Selection approach (Janes, Pepe, and Huang 2014) where they bootstrap the standard error and confidence intervals.

We will review each outcome and the specific calculations for each. We will generalize some of the approaches when able to for brevity and conciseness.

### 2.2.1 Generalized linear model

For this section we will discuss the generalized linear model and focus on a binary outcome and continuous outcome. For a binary outcome the model of choice is the logistic regression. We specify $Z$ as a continuous biomarker, $A = \{0,1\}$ is treatment, $Y = \{0,1\}$ is the response where we use logistic regression to fit the response to the covariates $W = (A, Z)$. The predicted risk is obtained from the logistic regression model. We assume $Y = 1$ is the worse outcome and the predicted risk is estimated with $P(Y|A,Z)$. If not then set the risk to be 1- $P(Y|A,Z)$ or flip the $Y$ values to have $Y = 1$ be the worst.

The predicted risk without additional covariates is

$$P(Y|A,Z) = logit^{-1}(\beta_0 + \beta_1 A + \beta_2 Z + \beta_3 A*Z)$$

$$= \frac{\exp(\beta_0 + \beta_1 A + \beta_2 Z + \beta_3 A*Z)}{1 + \exp(\beta_0 + \beta_1 A + \beta_2 Z + \beta_3 A*Z)},$$

and for each treatment arm is

$$P(Y|A=1,Z) = logit^{-1}(\beta_0 + \beta_1 + \beta_2 Z + \beta_3 Z)$$

$$P(Y|A=0, Z) = logit^{-1}(\beta_0 + \beta_2 Z).$$

The predicted risk for each treatment arm versus the biomarker is plotted to visualize the relationship of biomarker and outcome by treatment and see if the lines cross.

The predicted risk can also be obtained when there are covariates by extending the approach above. We specify $Z$ as a continuous biomarker, $A = \{0,1\}$ is treatment, $\boldsymbol{X}$ as other covariates such as prognostic factors, $Y = \{0,1\}$ is the response where we use logistic regression to fit the response to the covariates $W = (A, Z, \boldsymbol{X})$.

The predicted risk is $P(Y|A, Z, X) = logit^{-1}(\beta_0 + \beta_1 A + \beta_2 Z + \beta_3 A*Z + \beta_p^T \boldsymbol{X})$

$$= \frac{\exp(\beta_0 + \beta_1 A + \beta_2 Z + \beta_3 A*Z + \beta_p^T \boldsymbol{X})}{1 + \exp(\beta_0 + \beta_1 A + \beta_2 Z + \beta_3 A*Z + \beta_p^T \boldsymbol{X})},$$

and for each treatment arm is

$$P(Y|A=1, Z, X) = logit^{-1}(\beta_0 + \beta_1 + \beta_2 Z + \beta_3 Z + \beta_p^T \boldsymbol{X})$$

$$P(Y|A=0, Z, X) = logit^{-1}(\beta_0 + \beta_2 Z + \beta_p^T \boldsymbol{X}).$$

Since we need a probability of the outcome for biomarker and treatment arm, we need to marginalize the probability over the covariates. In order to marginalize over the covariates we sum over all the covariate values from the observations, but keep the treatment and biomarker value fixed; and do this for each biomarker value:

$$P(Y|A=1, Z) = \frac{\sum_{i=1}^{n} \frac{\exp(\beta_0 + \beta_1 + \beta_2 Z + \beta_3 Z + \beta_4 X_{4i} + .. + \beta_p X_{pi})}{1 + \exp(\beta_0 + \beta_1 + \beta_2 Z + \beta_3 Z + \beta_4 X_{4i} + \beta_p X_{pi})}}{n}$$

$$P(Y|A=0, Z) = \frac{\sum_{i=1}^{n} \frac{\exp(\beta_0 + \beta_2 Z + \beta_4 X_{4i} + .. + \beta_p X_{pi})}{1 + \exp(\beta_0 + \beta_2 Z + \beta_4 X_{4i} + \beta_p X_{pi})}}{n}.$$

We apply G-computation to obtain the marginalized risk. Once we have $(Y|A = 0, Z)$ and $P(Y|A = 1, Z)$ we can then plot the predicted lines and also find the cut-offs. We have a similar approach to the treatment selection approach. The predicted risk confidence interval (CI) is based on the standard error (SE) derived from Huber Sandwich Estimator which is robust to misspecification of the covariance stucture. This used avg_predictions and avg_comparisons function from the R marginaleffects package (Arel-Bundock 2025) to obtain the SE and CI of each treatment arm and its difference, respectively.

For a continuous outcome the model of choice is the linear regression. We specify $Z$ as a continuous biomarker, $A$={0,1} is treatment, and $Y$ is a continuous response where we use linear regression to fit the response to the covariates $W = (A, Z)$. The predicted risk is obtained from the linear regression model. We assume the highest value of $Y$ is the worse outcome and the predicted risk is estimated with $E(Y|A, Z)$. We can also standardize Y to then take the probability.

The predicted risk without additional covariates is $E(Y|A, Z) = \beta_0 + \beta_1 A + \beta_2 Z + \beta_3 A * Z$, and for each treatment arm is

$$E(Y|A = 1, Z) = \beta_0 + \beta_1 + \beta_2 Z + \beta_3 Z$$

$$E(Y|A = 0, Z) = \beta_0 + \beta_2 Z.$$

As demonstrated above, the predicted risk can also be obtained when there are covariates by extending the approach above. We specify $Z$ as a continuous biomarker, $A = \{0,1\}$ is treatment, $\boldsymbol{X}$ as other covariates such as prognostic factors, $Y = \{0,1\}$ is the response where we use linear regression to fit the response to the covariates $W = (A, Z, \boldsymbol{X})$.

The predicted risk is $E(Y|A, Z, X) = \beta_0 + \beta_1 A + \beta_2 Z + \beta_3 A * Z + \beta_p^T \boldsymbol{X}$, and for each treatment arm is

$$E(Y|A = 1, Z, X) = \beta_0 + \beta_1 + \beta_2 Z + \beta_3 Z + \beta_p^T \boldsymbol{X}$$

$$E(Y|A = 0, Z, X) = \beta_0 + \beta_2 Z + \beta_p^T \boldsymbol{X}.$$

Since we need an expected mean of the outcome for biomarker and treatment arm, we need to marginalize the mean over the covariates. In order to marginalize over the covariates we sum over all the covariate values from the observations, but keep the treatment and biomarker value fixed; and do this for each biomarker value:

$$E(Y|A=1, Z) = \frac{\sum_{i=1}^{n} \beta_0 + \beta_1 + \beta_2 Z + \beta_3 Z + \beta_4 X_{4i} + .. + \beta_p X_{pi}}{n}$$

$$E(Y|A=0, Z) = \frac{\sum_{i=1}^{n} \beta_0 + \beta_2 Z + \beta_4 X_{4i} + .. + \beta_p X_{pi}}{n}.$$

We apply G-computation to obtain the marginalized risk. Once we have $E(Y|A=0, Z)$ and $E(Y|A=1, Z)$ we can then plot the predicted lines and also find the cut-offs. We have a similar approach to the treatment selection approach (Janes, Pepe, and Huang 2014).

We have added CIs based on model-based estimates for binary and continuous outcome, using the Huber sandwich estimator, which is robust to misspecification of covariance structure. This used avg_predictions and avg_comparisons function from the R marginaleffects package (Arel-Bundock 2025) to get SE and CI of each treatment arm and its difference, respectively.

### 2.2.2 Time-to-event endpoint

For this section we focus on survival time data. We define the data to be $(Y, \delta)$, where $Y = min(T, C)$ is the time-to-event outcome, $T$ is the failure time, $C$ is the censored time, $\delta = I(T \leq C)$ is the indicator of censorship for time. We can model time-to-event data with a Cox proportional hazards model

$h(t|A, Z) = h_o(t) \exp(\beta_1 A + \beta_2 Z + \beta_3 A * Z)$

The cumulative hazard is:

$$H(t|A, Z) = \int_0^T h(t|A,Z) dt = \int_0^T -h_o(t) \exp(\beta_1 A + \beta_2 Z + \beta_3 A * Z) dt =$$

$$= \exp(\beta_1 A + \beta_2 Z + \beta_3 A * Z)) \int_0^T -h_o dt$$

$$= \exp(\beta_1 A + \beta_2 Z + \beta_3 A * Z))\lambda(T),$$

and survival is:

$$S(t|trt, Z) = exp\bigl(-H(t|A, Z)\bigr) = \exp(\lambda(T)exp(\beta_1 A + \beta_2 Z + \beta_3 A * Z)).$$

A landmark analysis is needed to be done to estimate the predicted risk, where a specific time point is selected and the median survival time is suggested, however this may be adjusted based on the clinical team's decision. Due to proportional hazards the cut-off will be time-independent.

The predicted risk is $P(t|A, Z) = 1 - S(t|A, Z)$

$$=1 - \exp(-\lambda\exp(\beta_1 A + \beta_2 Z + \beta_3 A * Z)),$$

and for each treatment arm

$$P(Y|A = 1, Z) = 1 - \exp(-\lambda\exp(\beta_1 + \beta_2 Z + \beta_3 Z))$$

$$P(Y|A = 0, Z) = 1 - \exp(-\lambda\exp(\beta_2 Z)).$$

When additional covariates $X$ are considered, we now have covariates $W = (A, Z, X)$, and the Cox model is

$$h(t|W) = h_o(t)\exp(\beta_1 A + \beta_2 Z + \beta_3 A * Z + \beta_p^T X)$$

and survival is:

$$S(t|A, Z) = \exp(-\lambda\exp(\beta_1 A + \beta_2 Z + \beta_3 A * Z + \beta_p^T X)).$$

The predicted risk is $P(Y|A, Z, X) = 1 - S(t|A, Z, X)$

$$=1 - \exp(-\lambda\exp(\beta_1 A + \beta_2 Z + \beta_3 A * Z + \beta_p^T X)).$$

The predicted risk for each treatment arm is

$$P(Y|A = 1, Z, \boldsymbol{X}) = 1 - \exp(-\lambda \exp(\beta_1 + \beta_2 Z + \beta_3 Z) + \beta_p^T \boldsymbol{X})$$

$$P(Y|A = 0, Z, \boldsymbol{X}) = 1 - \exp(-\lambda \exp(\beta_2 Z + \beta_p^T \boldsymbol{X})).$$

We marginalize the predicted risk for each biomarker value by averaging the predicted risk across all the covariate values and while keeping treatment and biomarker fixed:

$$P(Y|A = 1, Z, \boldsymbol{X}) = \frac{\sum_{i=1}^{n} 1 - \exp(-\lambda_1 \exp(\beta_1 + \beta_2 Z + \beta_3 Z) + \beta_p^T \boldsymbol{X})}{n}$$

$$P(Y|A = 0, Z, \boldsymbol{X}) = \frac{\sum_{i=1}^{n} \sum_{i=1}^{n} 1 - \exp(-\lambda_1 \exp(\beta_2 Z + \beta_p^T \boldsymbol{X}))}{n}.$$

For survival endpoints we get estimates from the ATE function of the R riskRegression package (Gerds, Ohlendorff, and Ozenn 2023), which provides model-based SE using the delta method.

## 2.3    Cut-off

Prior to identifying a cut-off, we suggest a hierarchical strategy to assess whether a biomarker is predictive. First, we fit a model, plot the predicted fit with confidence intervals, and evaluate the p-value from interaction of treatment and biomarker as continuous. The confidence intervals and p-value should complement each other, but the confidence interval may be preferred as evidence due to the amount of uncertainty shown. If the confidence intervals of the predicted risk curves from treatment and control arms do not overlap or the p-value is declared to be statistically significant then this suggests the presence of a predictive biomarker. If the risk curves cross then it suggests the biomarker might be a predictive biomarker, treatment benefit switches and a cut-off can be easily identified. When the lines do not cross the cut-off found can be outside a meaningful value so it should be checked if the cut-off is contained in $(min_Z, max_Z)$ where $min_Z$ is the minimum of $Z$ and $max_Z$ is the maximum of $Z$.

To determine a cut-off for the biomarker that finds the biomarker positive and negative group we suggest the following. We can solve for the cut-off "theoretically" with the coefficients estimates, or estimate the cut-off empirically using linear interpolation obtained from the predicted risk plots, or obtained as the unique root where the predicted risk difference is 0 (e.g. using uniroot from R). Note that the unique root value may not be contained in the dataset. It is important that the identified cut-off should be clinically meaningful.

When there are no covariates or we fix the covariate values we can use the formula-based cut-off $Z_{cut\_formula} = -\beta_1/\beta_3$, where $\beta_1$ is the main effects coefficient of treatment and $\beta_3$ is the interaction between treatment and the biomarker. We can show this by solving for $P(Y|A = 0, Z) - P(Y|A = 1, Z) = 0$ and when covariates those values are fixed and not marginalized. This was shown by Mboup et al (Mboup, Blanche, and Latouche 2020) and Janes et al (Janes, Pepe, and Huang 2014) for the unadjusted model. When we marginalize the covariates, the model-based formula provides an estimate of the cut-off; and it is often close to the empirical estimate via linear interpolation and the unique root when the lines cross.

We know the slope is found by $(y_2 - y_1)/(x_2 - x_1)$, with the general formula of $y = \beta_0 + \beta_x x$, and the cut-off can be estimated with this information. The cut-off using linear interpolation is found by setting the predicted risk of the control arm and treated arm to be the same where $\beta_{0,1} + \beta_{1,1} z_{cut\_linint}$ =$\beta_{0,0} + \beta_{1,0} z_{cut\_linint}$ where $\beta_{0,1}$ and $\beta_{1,1}$ are the intercept and slope for the treated arm and $\beta_{0,0}$ and $\beta_{1,0}$ are the intercept and slope for the control arm. The cut-off solution for linear interpolation is $Z_{cut\_linint}$ = $(\beta_{0,1} - \beta_{0,0})/ (\beta_{1,0} - \beta_{1,1})$. The cut-off found above will give us the predicted risk $pred_{cut\_linint}$ = $\beta_{0,1} + \beta_{1,1} z_{cut\_linint}$. For linear interpolation we use the marginalized risk of control and treated, $P(Y|A = 0, Z)$ & $P(Y|A = 1, Z)$, to find the cut-off $Z_{cut\_linint}$.

*Figure 1 Predicted risk plots*

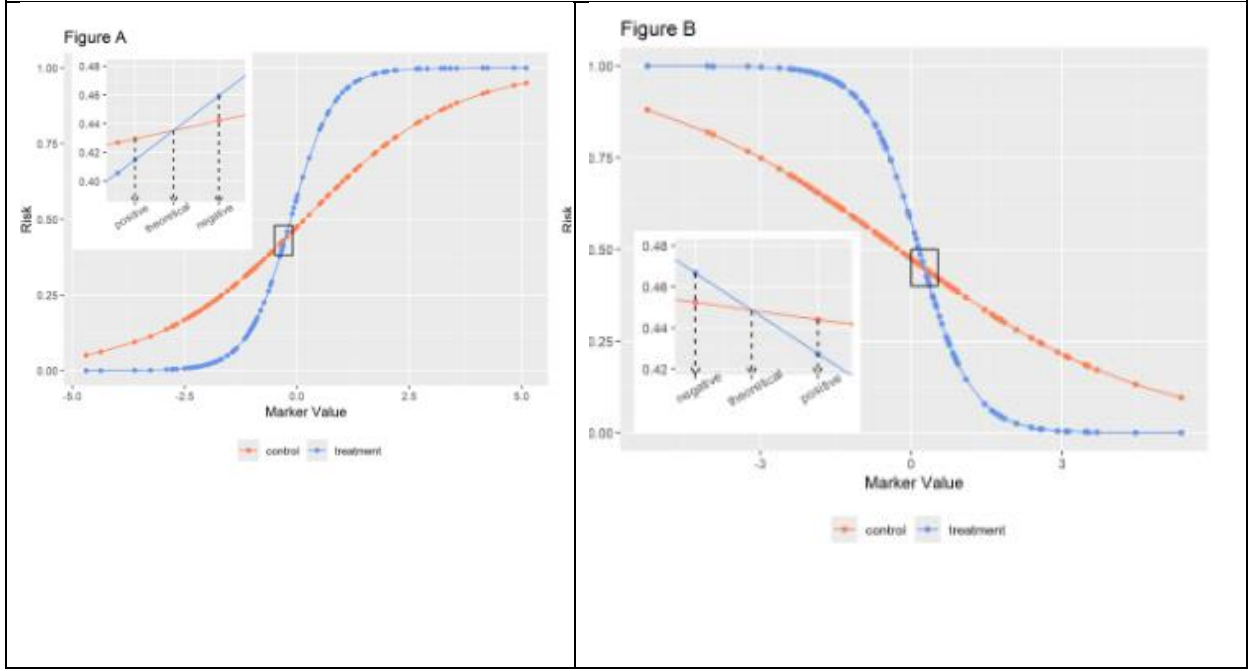

Three types of marker thresholds are provided from our PRIME package:

**The positive threshold** is the maximum marker value such that the risk in control is higher than the risk in treatment group if the risk difference (control - treatment) is positive for the minimum marker value (Figure 1A). Minimum marker value such that the risk in control is higher than the risk in treatment group if the risk difference (control - treatment) is negative for the minimum marker value (Figure 1B).

**The negative effect threshold** is the minimum marker value such that the risk in control is lower than the risk in treatment group if the risk difference (control - treatment) is positive for the minimum marker value (Figure 1A). Maximum marker value such that the risk in control is lower than the risk in treatment group if the risk difference (control - treatment) is negative for the minimum marker value (Figure 1B).

**The theoretical threshold** is the value where the risk difference is zero. We will determine this value from the predicted risk of a model through linear interpolation and should lie between the positive and negative effect thresholds (Figure 1A and 1B). Note that this value may not exist in the dataset.

## 2.4 Calibration assessment

Calibration assessment is an approach to assess goodness of fit and visualize it, and also important for marker evaluation. Model-based estimates will be used to inform the model fit.

Observations are split into $G$ evenly-size groups of size $n_g$ based on quantiles of predicted treatment effect $\hat{\Delta}(Z)$. For each group, the average predicted treatment effect $\overline{\Delta}_g(Z)$ can be calculated as $\frac{1}{n_g}\sum_{i \in g}[\hat{P}(Y|A = 0, Z_i) - \hat{P}(Y|A = 1, Z_i)]$ where the predicted risk $\hat{P}(Y|A = a, Z_i)$ is marginalized over covariates X.

When the covariates $X$ are categorical variables, the corresponding observed treatment effect in group $g$ is $\tilde{\Delta}_g(Z) = \frac{1}{n_g}\sum_{i \in g}\{E_X[\tilde{P}(Y|A = 0, Z_i, x) - \tilde{P}(Y|A = 0, Z_i, x)]\}$ where $\tilde{P}(Y|A = 0, Z_i, x) - \tilde{P}(Y|A = 0, Z_i, x)$ is the observed treatment effect conditioning on covariate value $x$ in group $g$. If a category $x$ does not appear in both treatment and control groups, it will be excluded from the calculation. For continuous covariate variables, they will be divided into $m$ groups based on the quantiles of the covariate. Subsequently, the observed treatment effect $\tilde{\Delta}_g(Z)$ can be calculated.

The goodness of fit is evaluated by plotting the $G$ observed treatment effects against the predicted treatment effects. Additionally, the observed treatment effects will be superimposed on the treatment effect curve.

## 2.5 Biomarker comparison and Net gain

Net gain is an approach to summarize the performance of a marker. This will facilitate a way to compare biomarkers. The following measures depend on the treatment rule from Janes et al (Janes et al. 2014), which we will extend to address covariates. Let the risk difference be $P(Y|A = 0, Z) - P(Y|A = 1, Z) = \Delta(Z)$.

We need to calculate the average benefit based on who is considered biomarker-positive and biomarker-negative. This can be done from the predicted risk curves with a fixed cut-off value. The net benefit measures require the following to estimate the summaries.

The average benefit of not receiving treatment among the biomarker-negative is:

$$B_{neg} = P(Y|A = 1, \Delta(Z) < 0) - P(Y|A = 0, \Delta(Z) < 0) = E(-\Delta(Z)|\Delta(Z) < 0).$$

The proportion of biomarker-negative is $P_{neg} = P(\Delta(Z) < 0)$.

A global measure of biomarker performance, denoted as $\Theta$, measures the decrease in the event rate under the biomarker-based treatment compared to the default 'treat all' strategy

$$\Theta = P(Y|A = 1) - [P(Y|A = 1, \Delta(Z) > 0)P(\Delta(Z) > 0) + P(Y|A = 0, \Delta(Z) < 0)P(\Delta(Z) < 0)]$$
$$= [P(Y|A = 1, \Delta(Z) < 0) - P(Y|A = 0, \Delta(Z) < 0)]P(\Delta(Z) > 0)$$
$$= B_{neg}P_{neg}.$$

Model based estimates will be used and the predicted risk is estimated from the model that marginalizes over the covariates. Bootstrapping can be used to estimate the SE and compare biomarkers. Biomarkers can be ranked by the net gain measure, with a higher value indicating a more informative biomarker and predictive capacity.

## 3 Other methods
### 3.1 Minimum p-value

The minimum p-value approach (Faraggi and Simon 1996; Woo and Kim 2020) will select the cut-off that corresponds to the minimum p-value of the interaction with treatment and biomarker. The GLM evaluated is:

$$f(y|A,Z,X) = g^{-1}\left(\beta_0 + \beta_1 A + \beta_2 I(Z>c) + \beta_3 A*I(Z>c) + \beta_p^T X\right) \text{ and}$$

the Cox model evaluated is

$$h(t|A,Z) = h_o(t)\exp\left(\beta_1 A + \beta_2 I(Z>c) + \beta_3 A*I(Z>c) + \beta_p^T X\right)$$

where $I$ is an indicator function. The model is evaluated at multiple $c$ cut-off values either selected from the actual data points or quantiles. The cut-off selected will be the one with the smallest p-value from the interaction between treatment and the dichotomized biomarker at cut-off $c$. Results will be shown unadjusted and adjusted by the Benjamini and Hochberg false discovery rate (FDR) p-value adjustment (Benjamini and Hochberg 1995).

### 3.2 Bayesian hierarchical model

A Bayesian hierarchical model (Chen, Jiang, and Tu 2014) has been developed for a time-to-event outcome under the framework of the Cox proportional hazards model. The Cox model is

$$h(t|A,Z) = h_o(t)\exp\left(\beta_1 A + \beta_2 I(Z>c) + \beta_3 A*I(Z>c) + +\beta_p^T X\right),$$

where $c$ is an unknown threshold parameter, and $Z$ is rescaled or normalized to (0,1) so that it represents the percentile of the population. A Bayesian approach is used for the statistical inference of the regression coefficients $\boldsymbol{\beta}$ and threshold parameter $c$. The probability density function (pdf) of the threshold parameter $c$ is $p_1(c|q) \propto q(q+1)c(1-c)^{q-1}$ where $q$ is a hyper-parameter with the density function $p_2(q) \propto \frac{(q-1)}{q(q+1)}$, $q > 1$. The regression coefficients $\boldsymbol{\beta}$ have a uniform improper prior distribution $p(\boldsymbol{\beta}) \propto 1$. A Markov Chain Monte Carlo (MCMC) approach and Gibbs Sampling is used to draw posterior samples from these marginal posterior distributions. Refer to Chen et al. (Chen, Jiang, and Tu 2014) for details of the steps for the MCMC algorithm for $(\boldsymbol{\beta}, c, q)$. For estimation of $c$, the average of the posterior MCMC samples are calculated with a credible interval for $c$.

Two approaches are offered for estimation and inference of $\boldsymbol{\beta}$:

1) The marginal approach: Use the posterior MCMC samples for a point estimate, to estimate the credible intervals, and to draw inference; and

2) The conditional approach: Use the estimate threshold value, $\hat{c}$, from the MCMC point estimate and given $\hat{c}$, estimate the $\boldsymbol{\beta}$ and its inference from the partial likelihood.

The R BHM package (Chen 2021) was used to estimate the cut-offs with this approach. The package includes other outcomes such as a binary outcome.

## 4 Simulation studies

We evaluate the operating characteristics of PRIME and compare it to the minimum p-value approach and Bayesian hierarchical model. A time-to-event endpoint is evaluated under various scenarios. The bias, standard deviation, $\sqrt{MSE}$, and coverages are assessed for the cut-off values. The mean of the net gain metric will also be calculated and compared.

The biomarker, $Z$, will be generated to be $Z \sim N(0.2,2)$; treatment, $A$, will be 50% of the sample; and two other covariates will be generated where $X_1 \sim N(0,1)$ and $X_2 \sim bin(n,1,0.5)$. The covariates are denoted $W = (A, Z, X_1, X_2)$. The survival time is $T \sim \exp(h)$ where $h = h_0 \exp(W^T \beta)$ and censoring time is $C \sim Unif(0, E)$. The survival time is $Y = \min(T, C)$ and event indicator is $\delta = I(T \leq C)$. It will be demonstrated when the biomarker has a strong and weak effect. To assess across the biomarker distribution, the true cut-off will be evaluated and fixed at 3 values 30%, 50%, 70% with values of (-0.85, 0.2, 1.25) – each simulation will have 1 cut-off. We denote the cut-off for the biomarker to be $Z_{cut}$. The fixed coefficient values are: $(\beta_Z, \beta_{X_1}, \beta_{X_2}) = (0.1, 0.1, 0.2)$, $(\beta_A, \beta_{A*Z}) = (0.5, 0.59)$ when $Z_{cut} = -0.85$, $(\beta_A, \beta_{A*Z}) = (-0.1, 0.5)$ when $Z_{cut} = 0.2$, and $(\beta_A, \beta_{A*Z}) = (-0.55, 0.44)$ when $Z_{cut} = 1.25$. Sample sizes of 200, 400, and 1000 are evaluated, for the 30% cut-off, median cut-off, 70% cut-off with a strong and weaker interaction effect (results in Supplement). We also evaluate power and type I error.

The simulation results are in Tables 1- 2 and Figures 1-2. Table 1 reports the bias, standard deviation, $\sqrt{MSE}$, coverage, and net gain for N=200 across the cut-offs. Figure 1 displays the absolute bias across the cut-offs and sample size. Figure 2 displays the $\sqrt{MSE}$ across the cut-offs and sample size. PRIME has the best operating characteristics with the smallest bias, smallest SD, smallest MSE, and most accurate coverages. The other approaches tend to be over-biased and select values where the control arm shows treatment benefit when the prevalence is below the upper quartiles (larger quartiles show under-bias). The BHM method does the second best where the operating characteristics tend be best with the median and do worse as the cut-off moves towards the tail [low/upper quartile]. The minimum p-value approach has extreme bias, large SD, large MSE, poor coverages and usually smallest net gain metric. The net gain is the largest with the PRIME approach, followed by the BHM approach; indicating the cut-off selected by PRIME shows the most treatment gain when using that cut-off. The plots show as the sample size increases the bias improves, MSE is smaller, and coverages are more accurate; where the results are stable across sample size and cut-offs with the PRIME approach and the minimum p-value and BHM approaches are more sensitive to the changes in the cut-off and the sample size. BHM performs well at the median and not as well at the tails as expected from previous findings from Chen et al (Chen, Jiang, and Tu 2014). The minimum p-value approach performs worse as the prevalence of the cut-off decreases. The findings are similar with a smaller interaction effect (shown in Supplement). Type I error and power are reported in Table 2. Prime has appropriate type-I error and power. The minimum p-value approach has inflated type-I error and is overpowered, indicating it will tend to select too many markers that will not be significant; whereas the FDR correction leads to an appropriate type-I error but is underpowered. BHM has slightly inflated type-I error rate and is underpowered but not as much as the minimum p-value with FDR correction.

| Table 1 Simulation results of strong biomarker predictive effect from survival outcome | | | |
|---|---|---|---|
| N=200, 20% censored | PRIME | Min p-value | BHM |
| True cut-off -0.85 | | | |
| Bias | -0.042 | 1.847 | 1.054 |
| SD | 0.387 | 1.044 | 0.714 |
| $\sqrt{MSE}$ | 0.389 | 2.121 | 1.273 |
| 95% coverage | 0.950 | 0.577 | 0.681 |
| Net gain | 0.235 | 0.162 | 0.204 |
| True cut-off 0.2 | | | |
| Bias | -0.011 | 0.952 | 0.107 |
| SD | 0.338 | 1.142 | 0.783 |
| $\sqrt{MSE}$ | 0.338 | 1.487 | 0.790 |
| 95% coverage | 0.951 | 0.879 | 0.950 |
| Net gain | 0.119 | 0.091 | 0.109 |
| True cut-off 1.25 | | | |
| Bias | 0.045 | -0.219 | -0.929 |
| SD | 0.444 | 1.217 | 0.812 |
| $\sqrt{MSE}$ | 0.447 | 1.236 | 1.233 |
| 95% coverage | 0.947 | 0.962 | 0.790 |
| Net gain | 0.051 | 0.034 | 0.030 |

| Table 2 Type I error and Power of $\beta_{A*Z}$ | | | | |
|---|---|---|---|---|
| N=200, 20% censored | PRIME | Min p-value | Min p-value FDR adj | BHM |
| True cut-off -0.85 | | | | |
| Type I error $(\beta_A, \beta_{A*Z}) = (0.20, 0)$ | 0.047 | 0.642 | 0.034 | 0.139 |
| Power, $(\beta_A, \beta_{A*Z}) = (0.20, 0.235)$ | 0.819 | 0.975 | 0.554 | 0.705 |
| True cut-off 0.2 | | | | |
| Type I error $(\beta_A, \beta_{A*Z}) = (-0.045, 0)$ | 0.053 | 0.639 | 0.052 | 0.128 |
| Power $(\beta_A, \beta_{A*Z}) = (-0.045, 0.225)$ | 0.802 | 0.979 | 0.541 | 0.685 |
| True cut-off 1.25 | | | | |
| Type I error $(\beta_A, \beta_{A*Z}) = (-0.286, 0)$ | 0.051 | 0.68 | 0.06 | 0.133 |
| Power $(\beta_A, \beta_{A*Z}) = (-0.286, 0.229)$ | 0.792 | 0.981 | 0.519 | 0.661 |

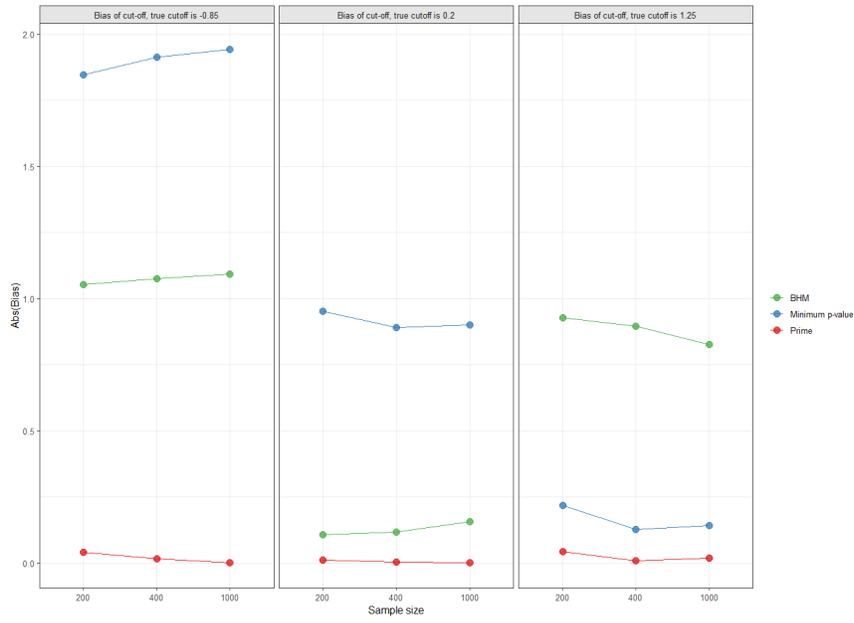

*Figure 1 Absolute bias across n and cut-offs*

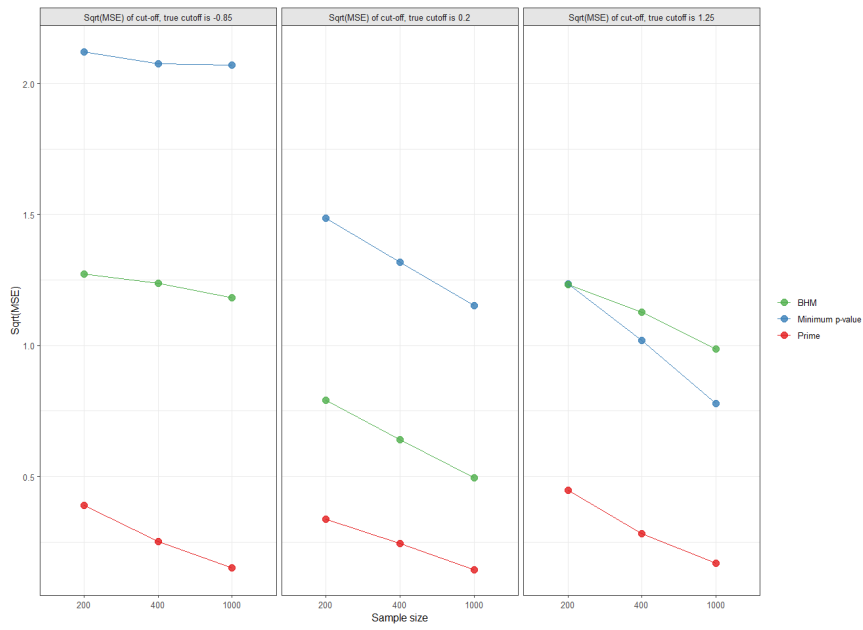

*Figure 2 Sqrt(MSE) across n and cut-offs*

## 5 Example

We will demonstrate the PRIME approach with a few examples. A meta-analysis study (Gurjao et al. 2020) [data provided at https://github.com/mirnylab/TMB_analysis] has multiple indications where subjects underwent immunotherapy. The focus of the study was to evaluate subjects taking Immune checkpoint blockade (ICB) treatments such as anti-CTLA-4 and anti-PD1, and combinations of therapy. These ICB treatments focus on regulatory pathways in T-lymphocytes to boost anti-tumor immune responses and have already demonstrated the ability to generate long-lasting clinical benefits for certain patients (Gurjao et al. 2020). We focused on 246 subjects with melanoma that were given combination therapy (n=118) or PD-1/PDL-1 therapy (n=128). Tumour Mutational Burden is a biomarker that measures total somatic nonsynonymous mutations in tumours in response to immune checkpoint blockade (ICB). This is partially because TMB is relatively simple to assess and measure. The biomarker of interest was tumor burden and we simulated circulating tumor DNA (ctDNA) data (see supplement for histograms of the data) and also added some noise to the original treatment data.

We adjusted for age with mean +/- SD (60.3 +/- 14.7) and gender (92 F/154 M), and genetic data was also available. The 2 outcomes of interest were overall survival and simulated ORR. The number of events was (47,43) with 41 months survival time for the combination therapy and PD-1/PDL-1 treatment, respectively. The number of responders was 69 (58%) and 76 (58%) for combination therapy and PD-1/PDL-1 treatment, respectively. ctDNA had non-overlapping predicted risk confidence intervals for overall survival and the interaction p-value of 0.0110 (statistically significant) (Figure 3b). The cut-off for ctDNA was 8.79 (51.2%) and the global performance measure was 0.031 under the treat all treatment rule. The minimum p-value approach found the cut-off to be 15.2 (31.3%) with a p-value of 0.0019 and FDR p-value adjusted of 0.092. The BHM approach also found a cut-off of 15.0 (32.5%) and p-value of 0.0042. TMB had overlapping predicted risk confidence intervals for overall survival and the interaction p-value of 0.355 (not statistically significant) (Figure 3b). The cut-off for TMB was 33.54 (82.1%), and the global performance measure was 0.010 under the treat all treatment rule. The minimum p-value found the cut-off to be 18.4 (73.6%) with a p-value of 0.0041 and FDR adjusted p-value adjusted of 0.075. The BHM approach also found a cut-off of 7.9 (44.7%) and p-value of 0.0140. Our analysis showed that ctDNA is a stronger predictive biomarker than TMB based on the risk plots and global marker performance measure. The calibration plots show ctDNA is a better fit than TMB. Although a trend is shown that ctDNA and TMB are predictive biomarkers with ORR, there is not strong evidence based on the predicted risk plots and summary measures. The minimum p-value and the BHM approaches did find cut-offs further along the tail of the distribution but that was in the right direction in showing treatment benefit. These values would select a smaller prevalence.

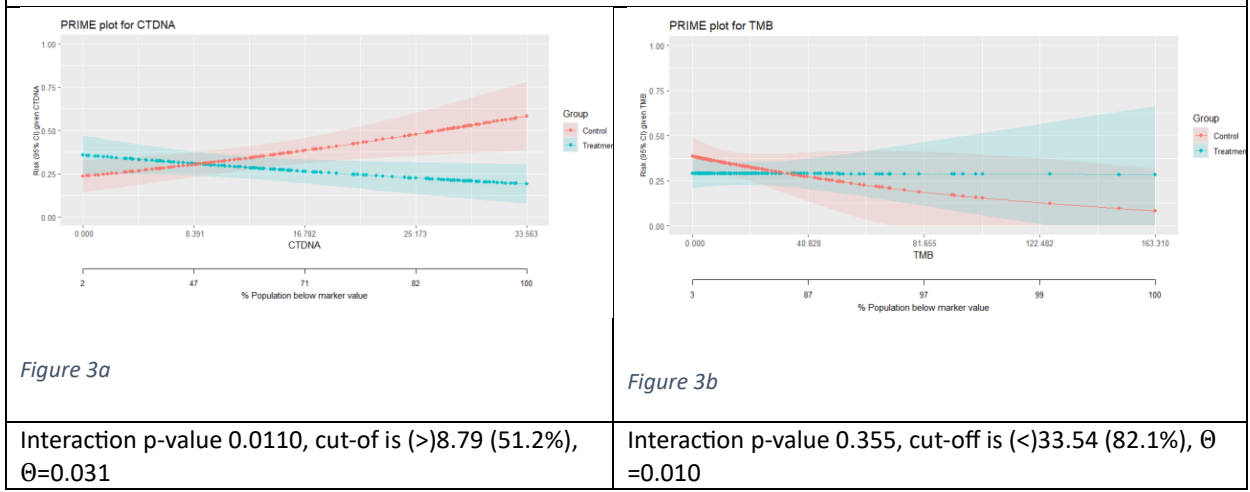

Figure 3 Predicted risk plots for the melanoma example with survival endpoint

| Figure 3a | Figure 3b |
|---|---|
| Interaction p-value 0.0110, cut-of is (>)8.79 (51.2%), Θ=0.031 | Interaction p-value 0.355, cut-off is (<)33.54 (82.1%), Θ=0.010 |

Figure 4 Treatment effect and calibration plots for the melanoma example with survival outcome

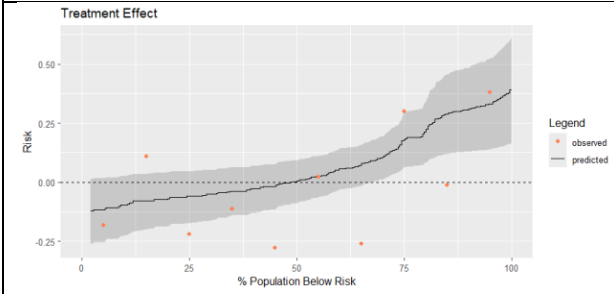

*Figure 4a Treatment effect plot for ctDNA*

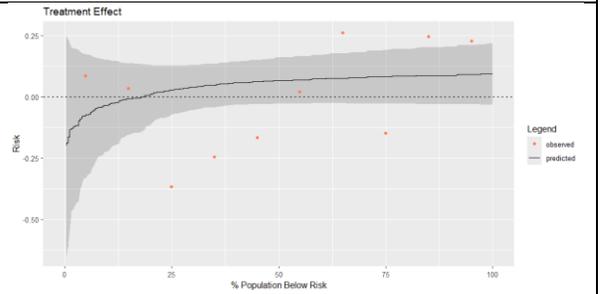

*Figure 4b Treatment effect plot for TMB*

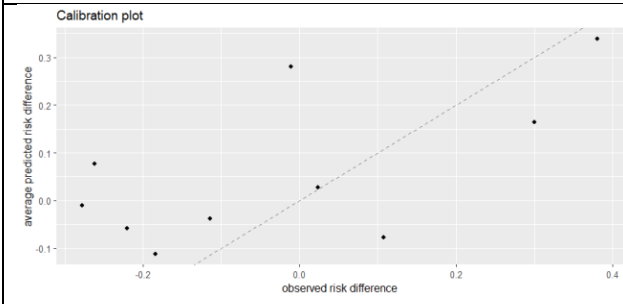

*Figure 4c Calibration plot for ctDNA*

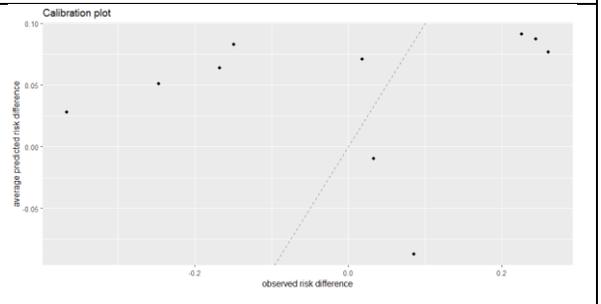

*Figure 4d Calibration plot for TMB*

| Figure 5 Predicted risk plots for the melanoma example with binary endpoint |
|---|

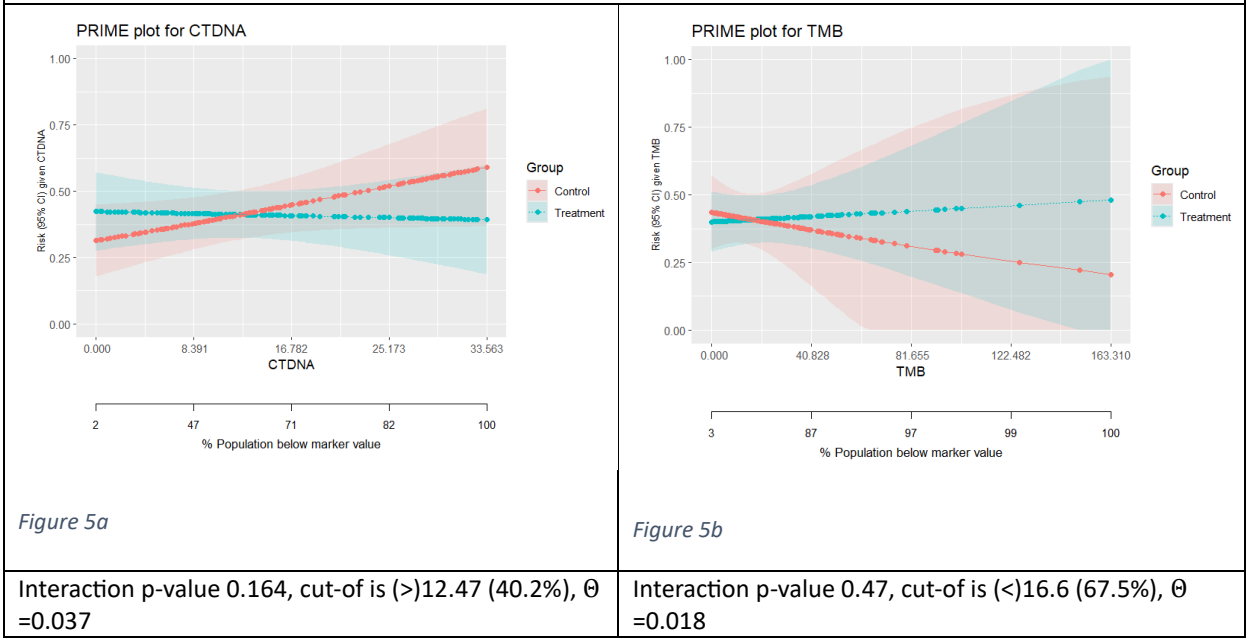

| *Figure 5a* | *Figure 5b* |
|---|---|
| Interaction p-value 0.164, cut-of is (>)12.47 (40.2%), Θ =0.037 | Interaction p-value 0.47, cut-of is (<)16.6 (67.5%), Θ =0.018 |

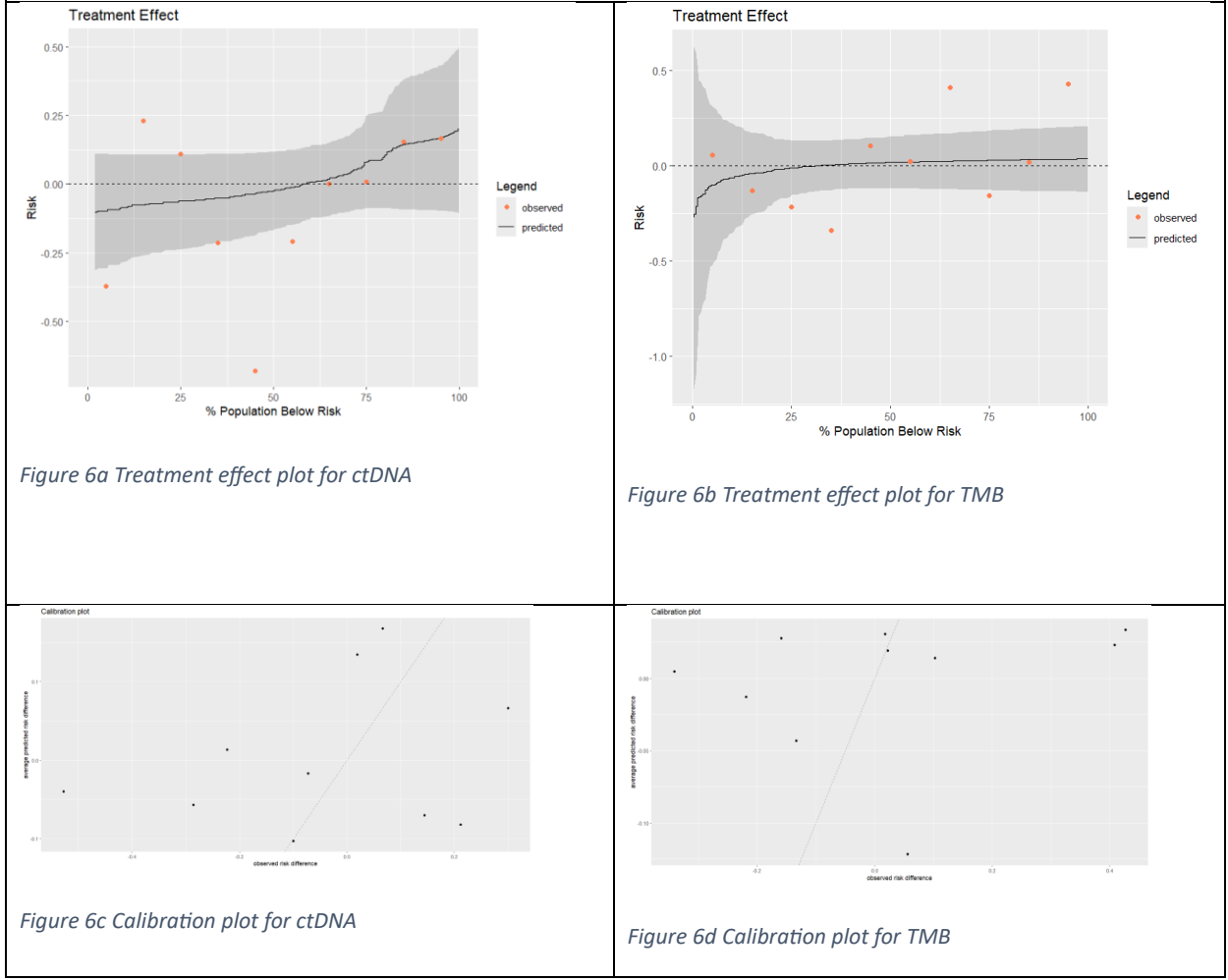

Figure 6 Treatment effect and calibration plots for the melanoma example with binary outcome

*Figure 6a Treatment effect plot for ctDNA*

*Figure 6b Treatment effect plot for TMB*

*Figure 6c Calibration plot for ctDNA*

*Figure 6d Calibration plot for TMB*

## 6  Discussion

Precision medicine is an important area to help advance the development of therapy for the right patients. Biomarkers help identify the right patient population. A predictive biomarker is defined as a biomarker that indicates how a patient is likely to respond to a specific treatment, signifying that the relationship between treatment and outcome varies across different biomarker subgroups. These biomarkers are highly sought after in our industry due to their potential to identify patient populations that are more likely to respond favorably to treatment.

Predictive biomarkers display varying treatment responses, and the connection between efficacy and treatment is influenced by the biomarker group. To evaluate the predictive qualities of a biomarker, a randomized controlled trial must include both a control group and a treatment group. In trials that feature only a treatment group, one can only assert the presence of a prognostic biomarker, as it is not possible to determine whether the relationship between efficacy and treatment varies by biomarker subgroup.

While both prognostic and predictive biomarkers help identify patient subgroups that respond better to treatments, understanding how responses differ requires the availability of data from multiple treatment options. We can ascertain whether a particular subgroup is more responsive to one treatment over another only when data from both treatments is analyzed. It is noteworthy that establishing cut-off points for prognostic biomarkers is typically more straightforward than for predictive biomarkers. With prognostic biomarkers, the ground truth is evident since each subject has a known and recorded outcome value. In contrast, determining predictive cut-offs is less direct because the individual treatment effect (ITE) needed for predictive biomarkers is not accessible through ROC curves. Therefore, custom methodologies are necessary to establish cut-off points for predictive biomarkers. The challenges surrounding predictive aspects are due to: 1) fundamental issues such as the unobservable nature of ITE and the requirement for causal inference and interaction effects, and 2) the fact that trials are often inadequately sized to meet the objectives of predictivity and test interaction terms (Gehlman [59]).

We demonstrated our PRIME approach to identify if a biomarker is predictive, find the cut-off, and compare biomarkers. PRIME was compared to the minimum p-value and Bayesian hierarchical model. The methods were compared with simulation studies to demonstrate operating characteristics. PRIME had the best operating characteristics with the smallest bias, smallest SD, smallest MSE, most accurate coverages, and appropriate type-I error and power. The other approaches tend to be over-biased and select values where the control arm shows treatment benefit when the prevalence is below the upper quartiles. The BHM method does the second best where the operating characteristics tend be best with the median and do worse as the cut-off moves towards the tail [low/upper quartile]. The minimum p-value approach has extreme bias, large SD, large MSE, poor coverages and usually smallest net gain metric. The net gain is the largest with the PRIME approach, followed by the BHM approach; indicating the cut-off selected by PRIME shows the most treatment gain when using that cut-off. The minimum p-value approach has inflated type-I error and is overpowered, whereas BHM has slightly inflated type-I error rate and is underpowered but not as much as the minimum p-value with FDR correction.

The PRIME approach was demonstrated with a melanoma cancer study and breast cancer study. We showed how to compare biomarkers and determine which one is likely to be predictive and find the population likely to respond to treatment. We also demonstrated the minimum p-value and BHM approaches.

Some future directions will be to incorporate bootstrapping to compare the biomarkers. We are also developing predictive approaches to address nonproportional hazards with flexible parametric models. Extension work for the Bayesian hierarchical model is underway to address the modelling of the biomarker and incorporate historical data to inform the priors. The Cox model work is being extended to evaluate single arm trials to make informed decisions for Phase 3 trials with historical and external data.

Precision medicine depends on biomarkers to find the right patient population for patient enrichment. We suggest our PRIME approach to guide decision on biomarkers in precision medicine provided the appropriate data is available. We have a beta version of the PRIME package and can share per request.

# Supplement: Predictive biomarker graphical approach (PRIME) for Precision medicine

Gina D'Angelo, Xiaowen Tian, Chuyu Deng, Xian Zhou

## Variances of betas

### GLM

This used avg_predictions and Robust Sandwich Covariate Estimate to get SE and CI from the R ATE package. The covariance estimate is:

$$Var(\beta) = (W^T W)^{-1} W^T \Omega W (W^T W)^{-1}, \Omega = \sigma^2 I ,$$

where the robust standard error $W^T \Omega W$ will be replaced by another estimate, $\frac{\mu_i^2}{(1-h_i)^2}$, based on the hat matrix used instead where $\mu_i$ the residuals and $h_i$ are from the hat matrix.

### Cox proportionals hazard

The Q-model is the model of interest and the ATE is the average over the individual and their covariate values while we fix treatment and biomarker value.

$$ATE_i = E[Y \mid A = 1, z_i, X] - E[Y \mid A = 0, z_i, X]$$

$$\frac{\partial}{\partial \beta} ATE_i = \frac{\partial}{\partial \beta}\left(P(Y \mid A = 0, Z) - P(Y \mid A = 1, Z)\right)$$

$$\text{Se(ATE)} = \begin{bmatrix} \frac{\partial}{\partial \beta_1} ATE & \frac{\partial}{\partial \beta_3} ATE \end{bmatrix} \begin{bmatrix} \sigma^2_{\beta_1} & \sigma_{\beta_1,\beta_3} \\ \sigma_{\beta_1,\beta_3} & \sigma^2_{\beta_3} \end{bmatrix} \begin{bmatrix} \frac{\partial}{\partial \beta_1} ATE \\ \frac{\partial}{\partial \beta_3} ATE \end{bmatrix} \text{ add for rest of } \partial \beta$$

$$\text{SE(ATE)} = \frac{\partial}{\partial \beta_1} ATE \, \sigma^2_{\beta_1} \, \frac{\partial}{\partial \beta_1} ATE$$

## Linear interpolation cut-off

We know the slope is found by $(y_2 - y_1)/(x_2 - x_1)$, with the general formula of y=intercept+slope*x, and Intercept=y-slope*x. For the control arm: a) r1 is slope of control where r1=(y2-y1)/(x2-x1), y is the predicted risk of marker value 2 and marker value 1 and x are the biomarker values from marker value 2 and 1, and b) r0 is the intercept, where r0=y1-r1*x1, y1 is the predicted risk of marker value 1. For the treated arm: a) b1 is the slope for the treated and b0 is intercept for the treated b1= (y2-y1)/(x2-x1), b) b0 is the intercept and b0=y1-b1*x1 .

We set the predicted values of the treated and control arms to be equal pred2=pred1; & pred2= b0+b1*$Z_{cut\_linint}$ & pred1= r0+r1*$Z_{cut\_linint}$, so therefore

b0+b1*$Z_{cut\_linint}$=r0+r1*$Z_{cut\_linint}$

$Z_{cut\_linint}$ = (b0-r0)/(r1-b1)

The cut-off found above will give us the predicted risk below

$pred_{cut\_linint}$ = b0 + b1*$Z_{cut\_linint}$

## Simulation results

The biomarker, $Z$, will be generated to be $Z \sim N(0.2,2)$, treatment, $A$, will be 50% of the sample, and 2 other covariates will be generated where $X_1 \sim N(0,1)$ and $X_2 \sim bin(n, 1, 0.5)$. The covariates are denoted $W = (A, Z, X_1, X_2)$. The survival time is $T \sim \exp(h)$ where $h = h_0 \exp(W^T \beta)$ and censoring time is $C \sim Unif(0, E)$. The survival time is $Y = \min(T, C)$ and event indicator is $\delta = I(T \leq C)$; and we will have 20% censoring. It will be demonstrated when the biomarker has a strong and weak effect. To assess across the biomarker distribution, the true cut-off will be evaluated and fixed at 3 values 30%, 50%, 70% with values of (-0.85, 0.2, 1.25) – each simulation will have 1 cut-off. We denote the cut-off for the biomarker to be $Z_{cut}$. The fixed coefficient values are: $(\beta_Z, \beta_{X_1}, \beta_{X_2}) = (0.1, 0.1, 0.2)$, $(\beta_A, \beta_{A*Z}) = (0.5, 0.59)$ when $Z_{cut} = -0.85$, $(\beta_A, \beta_{A*Z}) = (-0.1, 0.5)$ when $Z_{cut} = 0.2$, and $(\beta_A, \beta_{A*Z}) = (-0.55, 0.44.)$ when $Z_{cut} = 1.25$. Sample sizes of 200, 400, and 1000 are evaluated, for the 30% cut-off, median cut-off, 70% cut-off with a strong and weaker interaction effect. We also evaluate power and type I error. When there is a weak interaction effect the fixed coefficient values are: $(\beta_Z, \beta_{X_1}, \beta_{X_2}) = (0.1, 0.1, 0.2)$, $(\beta_A, \beta_{A*Z}) = (0.25, 0.29)$ when $Z_{cut} = -0.85$, $(\beta_A, \beta_{A*Z}) = (-0.05, 0.25)$ when $Z_{cut} = 0.2$, and $(\beta_A, \beta_{A*Z}) = (-0.275, 0.22)$ when $Z_{cut} = 1.25$.

Expected risk plots for each scenario mentioned are displayed to show the expected relationship between the outcome and biomarker by treatment arm. Plots S1-S3 is for the strong interaction effect.

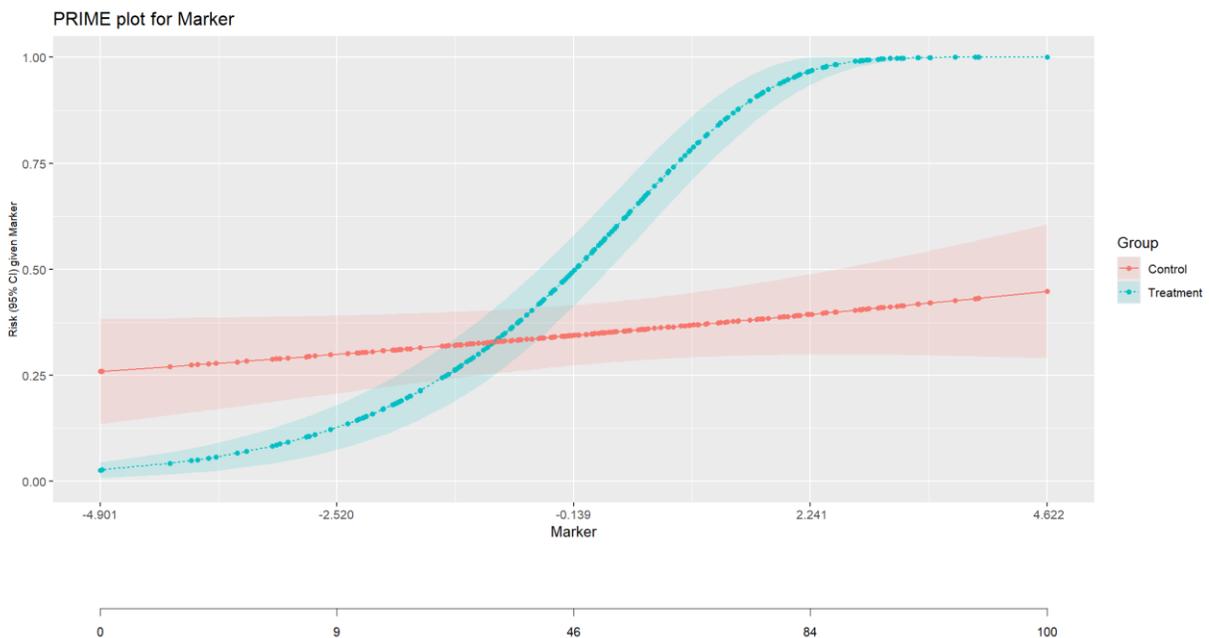

*Figure S7 Risk plot of survival vs biomarker with strong interaction effect and 30% cut-off*

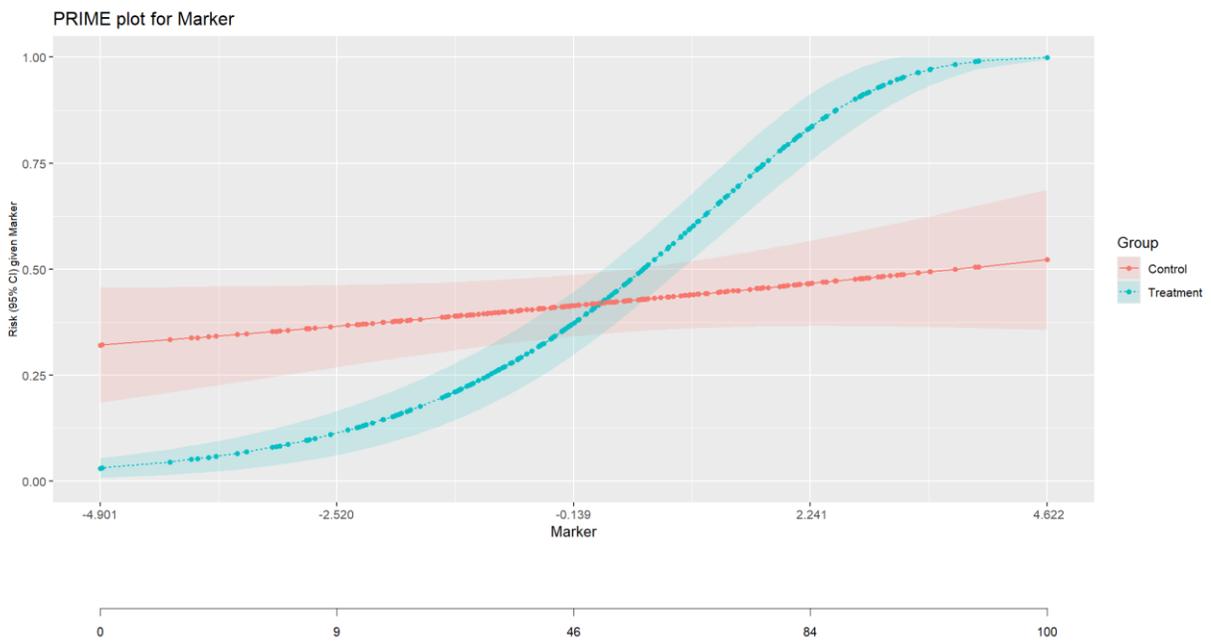

*Figure S8 Risk plot of survival vs biomarker with strong interaction effect and 50% cut-off*

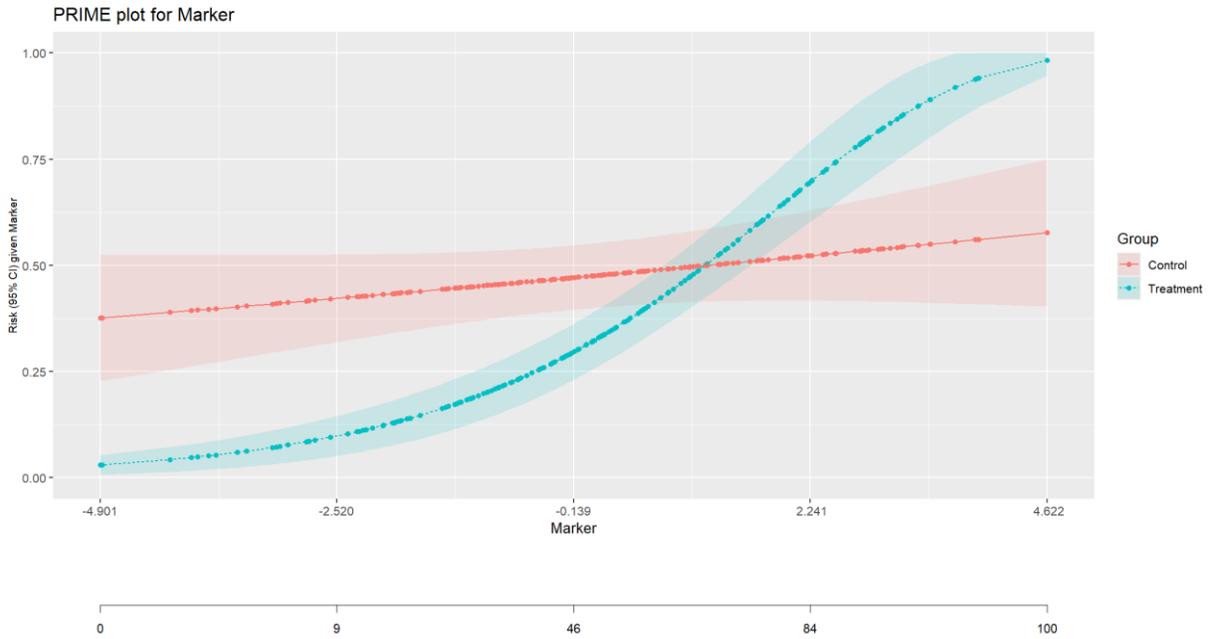

*Figure S9 Risk plot of survival vs biomarker with strong interaction effect and 70% cut-off*

Operating characteristics of the simulation studies for the strong interaction effect are displayed below for the absolute bias, standard deviation, sqrt MSE, coverages, and are assessed for the cut-off values (Figures S4-S8).

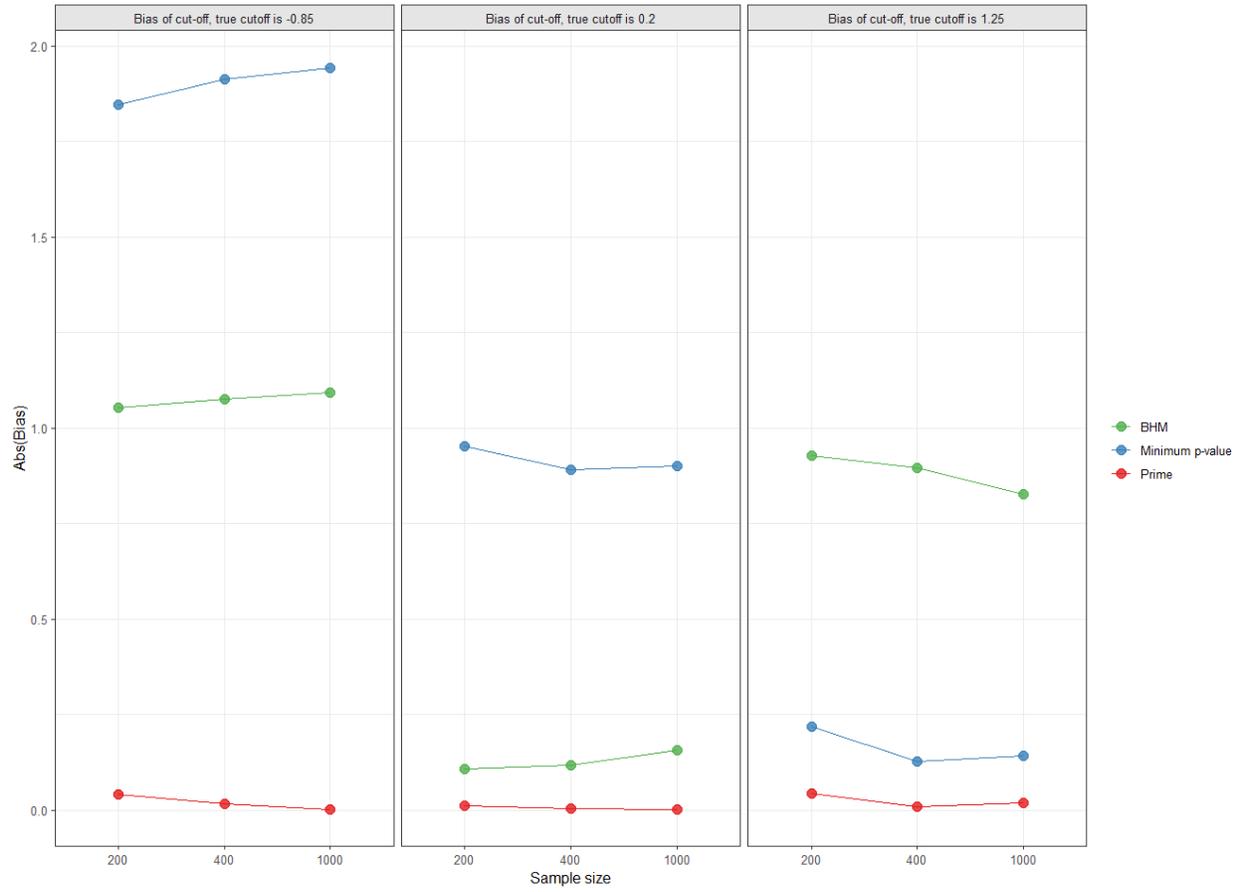

*Figure S4 Absolute bias across n and cut-offs for strong interaction effect*

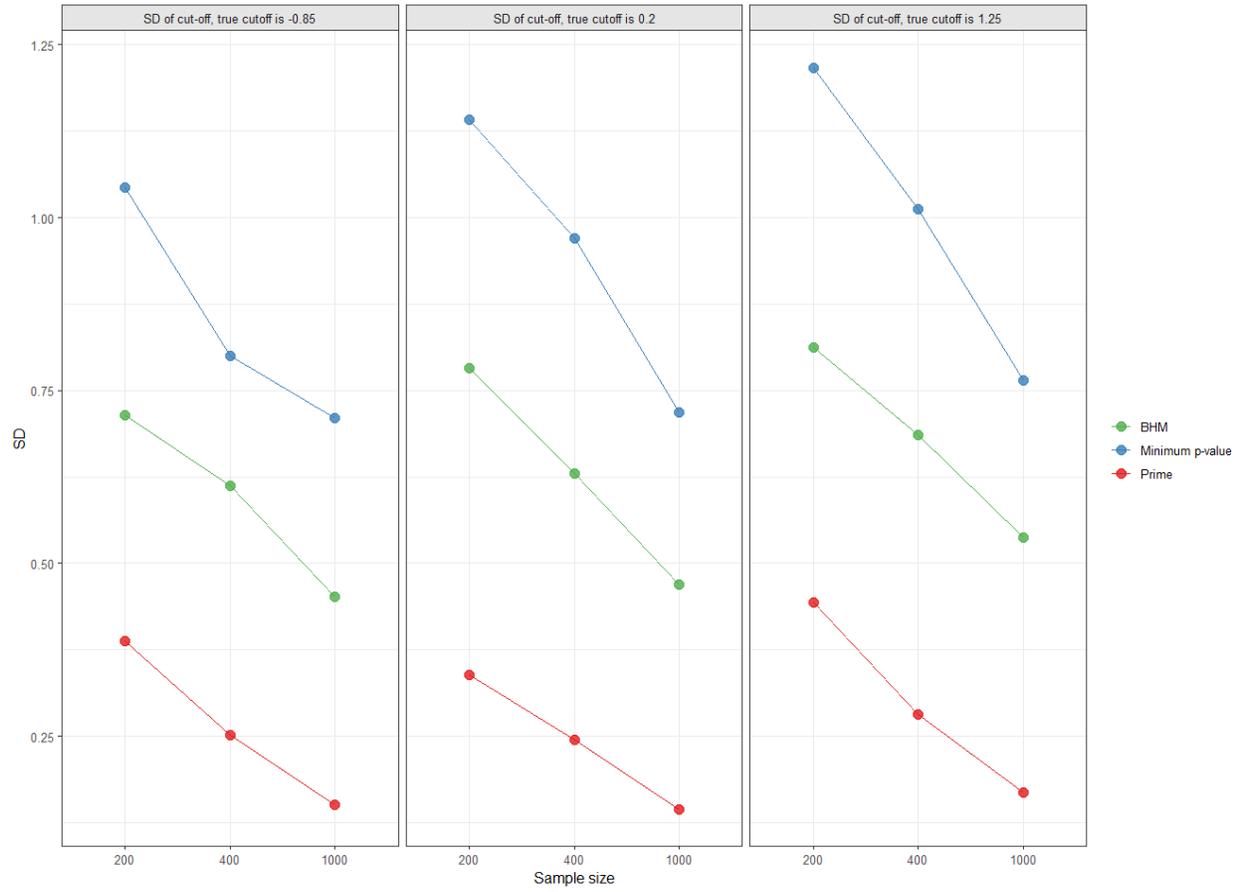

*Figure S5 SD across n and cut-offs for strong interaction effect*

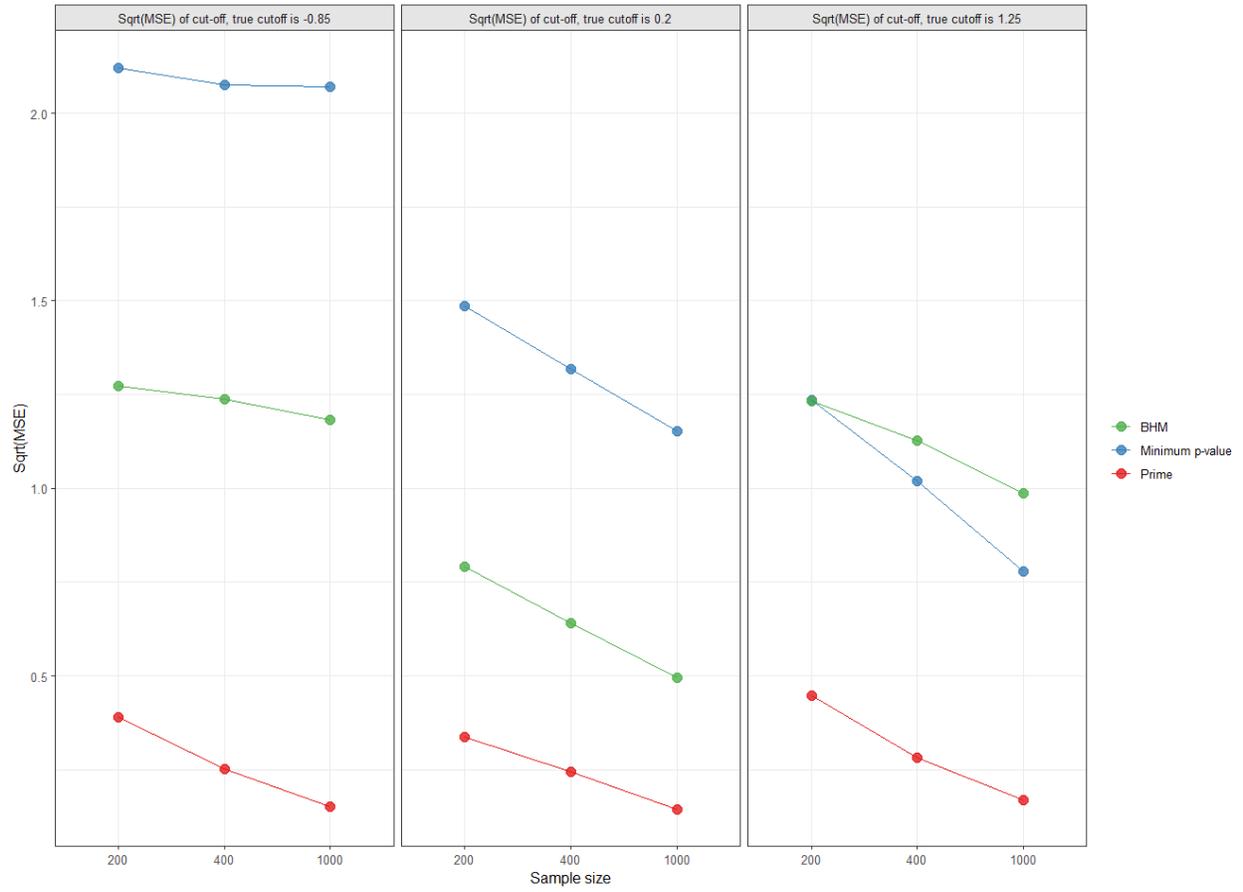

*Figure S6 Sqrt(MSE) across n and cut-offs for strong interaction effect*

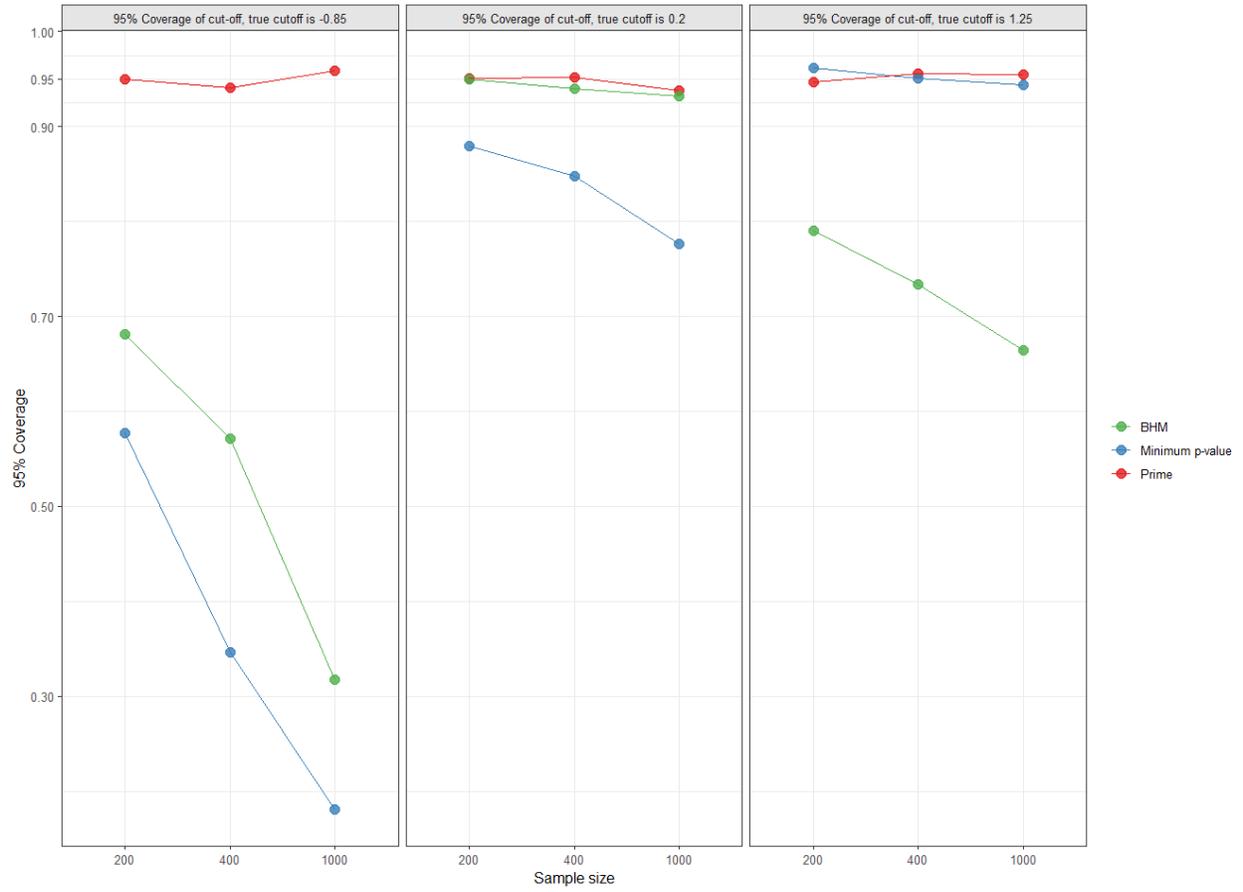

*Figure S7 Coverages across n and cut-offs for strong interaction effect*

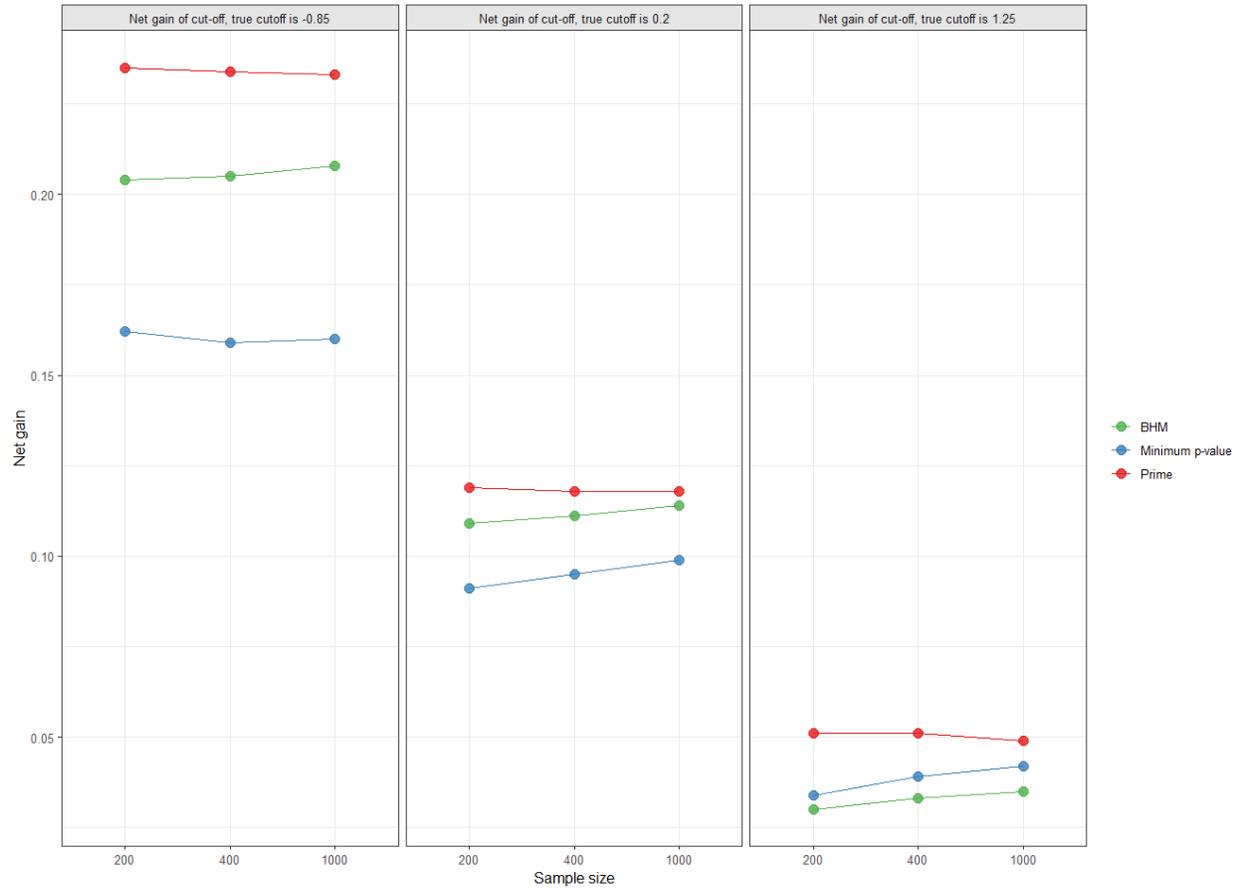

*Figure S8  Net gain across n and cut-offs for strong interaction effect*

Operating characteristics of the simulation studies for the strong interaction effect are displayed below for the absolute bias, standard deviation, sqrt MSE, coverages, and are assessed for the cut-off values (Figures S9-S13).

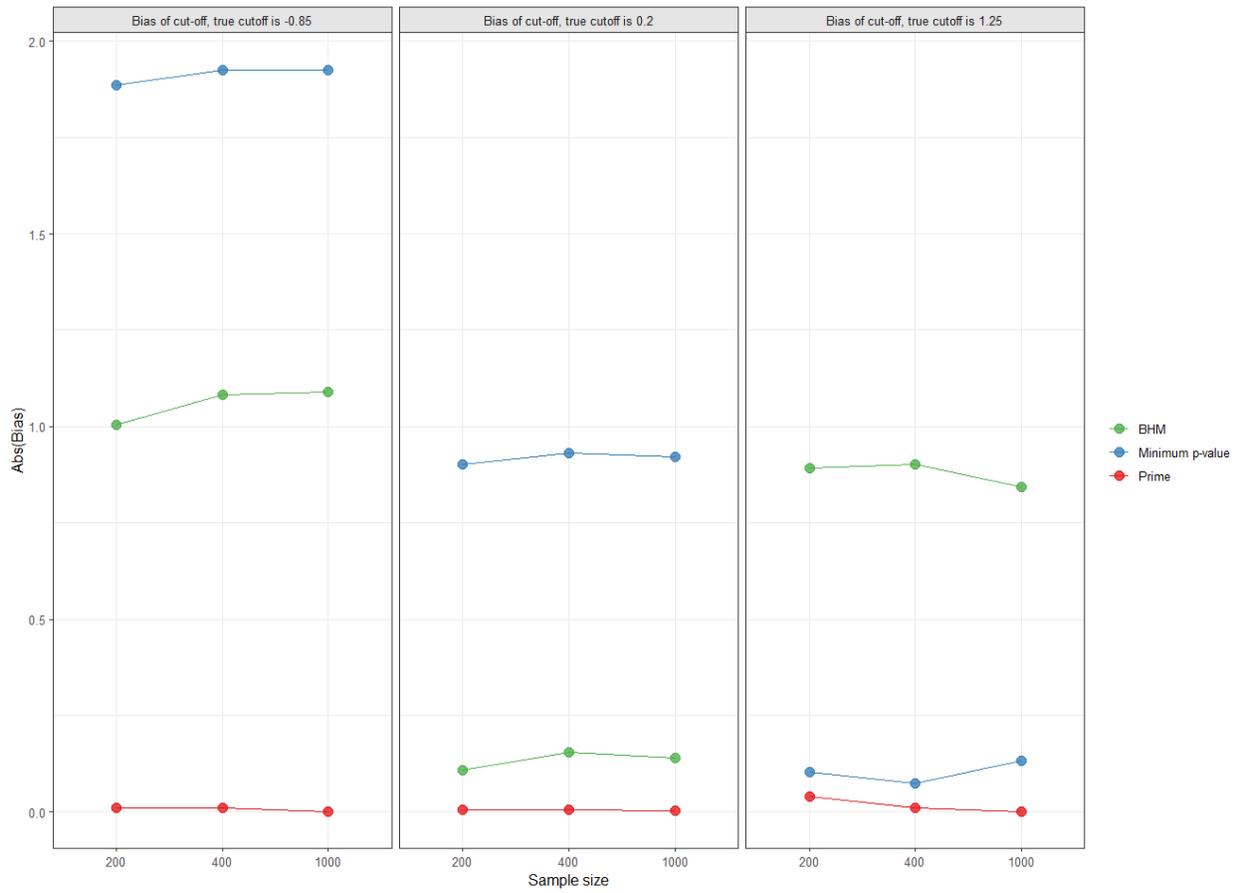

*Figure S9 Absolute bias across n and cut-offs for weak interaction effect*

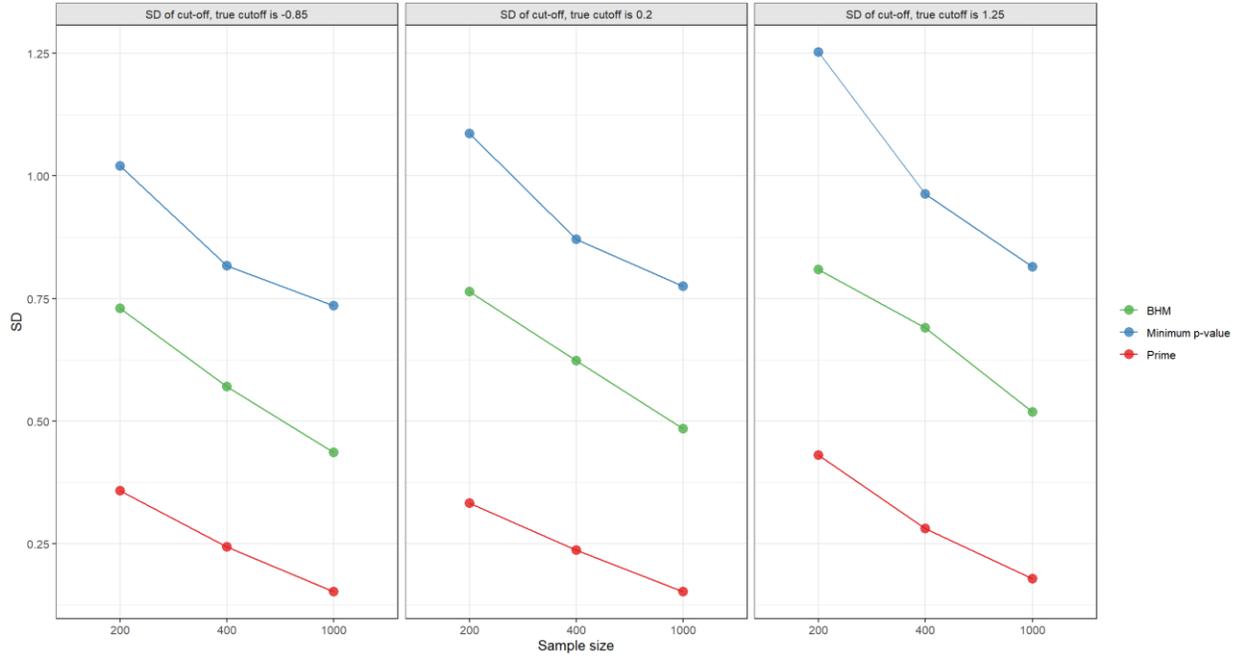

*Figure S10 SD across n and cut-offs for weak interaction effect*

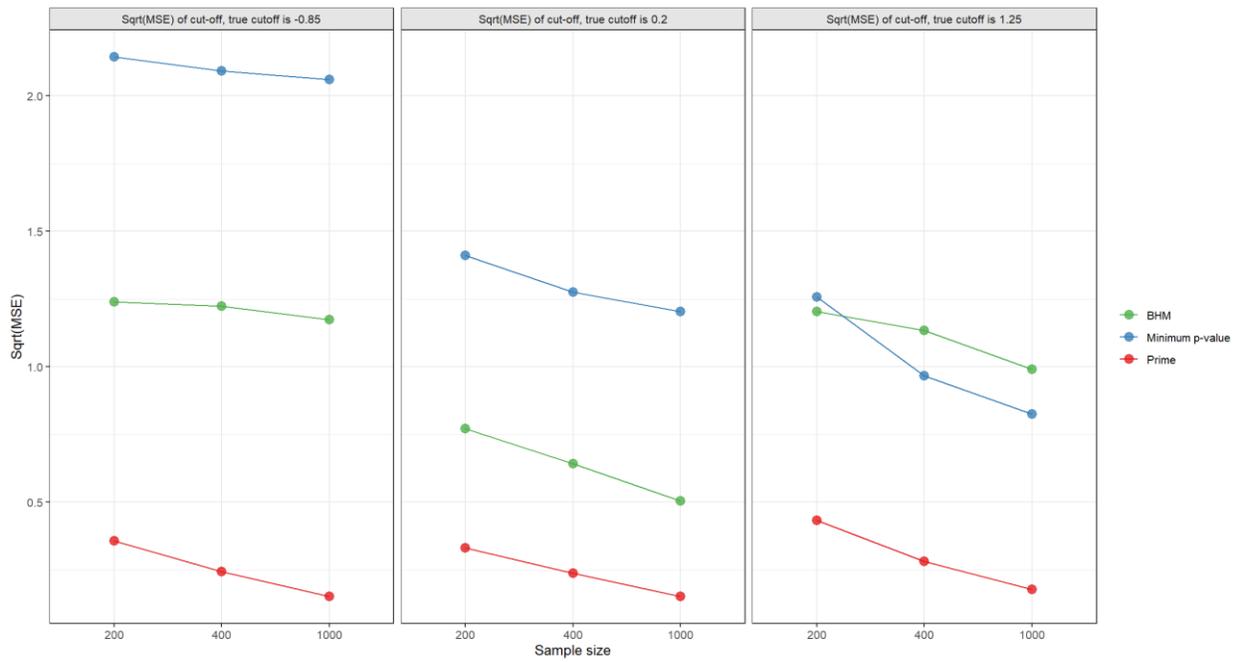

*Figure S11 Sqrt(MSE) across n and cut-offs for weak interaction effect*

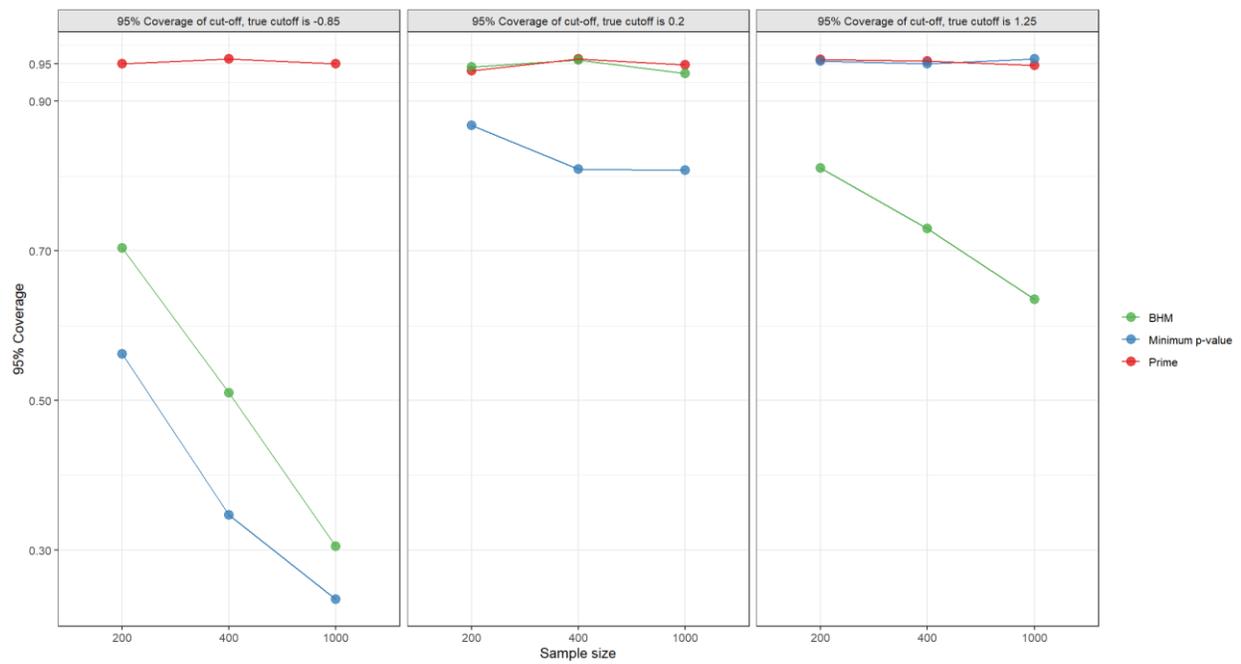

*Figure S12 Coverages across n and cut-offs for weak interaction effect*

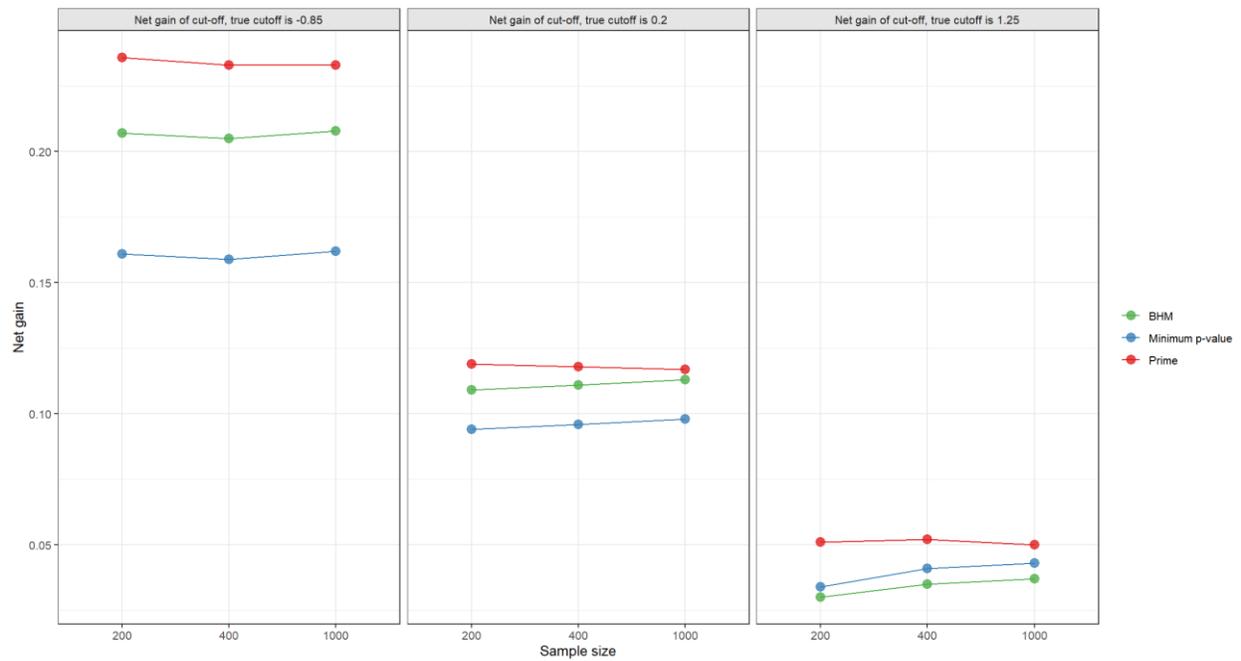

*Figure S13 Net gain across n and cut-offs for weak interaction effect*

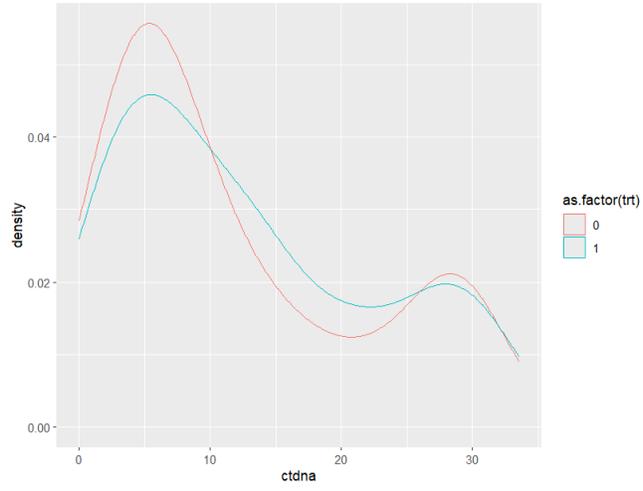

*Figure S14 Density plot of CTDNA*

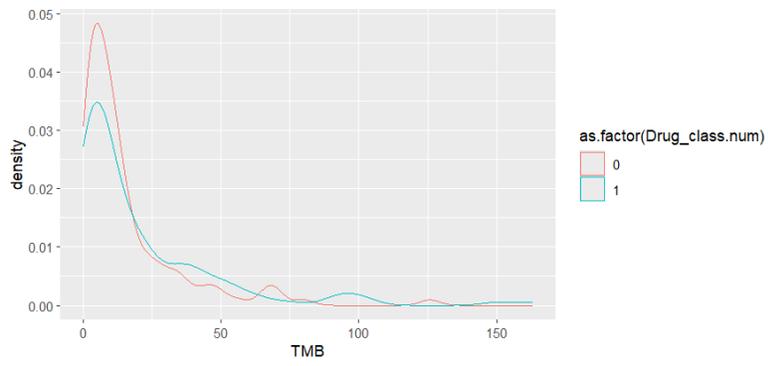

*Figure S15 Density plot of TMB*